\documentclass[%
 reprint,
 superscriptaddress,
 amsmath,amssymb,
 aps, prx,
]{revtex4-2}

\usepackage{graphicx}% Include figure files
\usepackage{dcolumn}% Align table columns on decimal point
\usepackage{bm}% bold math
\usepackage{xcolor}
\usepackage{booktabs}
\usepackage{hyperref}% add hypertext capabilities
\begin{document}
\pdfpageattr{/Rotate 0}

\author{Niclas Krupp}
\email{niclas.krupp@pci.uni-heidelberg.de}
\affiliation{%
 Theoretische Chemie, Physikalisch-Chemisches Institut, Universität Heidelberg, INF 229, 69120 Heidelberg, Germany 
}%
\author{Moritz Huber}
\author{Cheng Luo}
\author{Oriol Vendrell}
\email{oriol.vendrell@uni-heidelberg.de}
\affiliation{%
 Theoretische Chemie, Physikalisch-Chemisches Institut, Universität Heidelberg, INF 229, 69120 Heidelberg, Germany 
}%

%%%%%%%%%%%%%%%%%%%%%%%%%%%%%%%%%%%%%%%%%%%%%%%%%%%%%%%%%%%%%%%%%%%%%
\title
  {First principles simulation of the collective rovibronic ground state in a cavity}

\begin{abstract}
    Strong light–matter coupling in Fabry–Perot cavities can modify ground-state molecular reactivity, charge and energy transport, while modifications to single-molecule properties have not been observed experimentally. The mechanisms and reproducibility of such effects remain contested, with conflicting theoretical predictions driven by differences in Hamiltonian choice and quantum state representation. Here, we resolve these ambiguities with numerically exact quantum simulations of cavity-coupled molecular ensembles based on the \emph{ab initio} light–matter Hamiltonian, treating electrons, nuclei, and cavity photons on equal footing. We investigate ensembles of the rotational-vibrational-electronic Shin-Metiu model using variational tree-tensor-network quantum dynamics, capturing rovibronic couplings and anharmonicity. Embedding ensembles in a cavity induces local modifications of rotational, nuclear, and—more weakly—electronic observables in individual molecules. The extent of these modifications depends only on the per-molecule coupling strength up until \emph{each molecule} reaches the ultrastrong coupling regime, which remains unattainable in practical Fabry–Perot setups. Increasing ensemble size toward the thermodynamic limit causes these local modifications to vanish regardless of dipole self-energy inclusion. Nonetheless, global ground-state observables, such as light–matter coupling energy contributions (affecting overall polarizability) and cavity field displacement fluctuations, depend crucially on proper treatment of intermolecular and light–matter correlations, whereas intramolecular observables remain largely insensitive. These insights are crucial for guiding future investigations using approximate quantum treatments, and for the interpretation of experimental results in polaritonic chemistry.
\end{abstract}

\maketitle
\section{Introduction}
Strong coupling between reactive molecular systems and confined electromagnetic modes can
can result in modified chemical reactivity, both for thermal and photo-activated reactions, thereby
opening up pathways to mode-selective and tunable chemistry \cite{ebbesen2016hybrid,garcia2021manipulating,hirai2023molecular}. Depending on the cavity mode frequency, distinct properties and reactivities can be addressed: In the electronic strong coupling regime (ESC), cavity-enhanced selectivity of photochemical reactions \cite{hutchison2012modifying} and modified intersystem crossing dynamics \cite{yu2021barrier} could be demonstrated, among others \cite{xiang2024molecular}. Experiments in the vibrational strong coupling (VSC) regime have reported cavity-modification of electrical  \cite{kumar2024extraordinary} and ionic conductivity  \cite{fukushima2022inherent}, non-covalent interactions  \cite{joseph2021supramolecular}, as well as reaction rates \cite{lather2019cavity,thomas2019tilting,ahn2023modification}.
Cavity-modified reaction rates are particularly intriguing, as the prospect of controlling chemical reactivity with light has been a long-standing goal since the advent of lasers \cite{bloembergen1984energy,zewail1980laser}. Nonetheless, the extent to which cavities influence reactivity -- especially whether modifications of single-molecule properties are required for rate changes to occur --
remains actively debated \cite{thomas2024non,schwartz2025comment,schwennicke2024molecular}.
This ongoing uncertainty makes it challenging to establish clear and general conditions under which
cavity-controlled reactions could become
feasible \cite{fidler2023ultrafast,chen2024exploring,wang2021roadmap}.

Since the pioneering experiments by the group of Ebbesen \cite{thomas2016ground,thomas2019tilting}, several mechanistic explanations have been put forward, including alteration of the potential energy barrier due to strong coupling \cite{galego2019cavity,riso2022molecular}, new dynamic energy redistribution pathways among molecules and within molecules \cite{sun2022suppression,sun2023modification,li2021collective}, cavity-induced polarization hotspots \cite{sidler2024unraveling}, and optical filtering effects \cite{schwennicke2024molecular}. Experimental verification of these proposed mechanisms in the collective coupling regime is still outstanding \cite{campos2023swinging}. This is aggravated by the lack of reproducibility of some experiments \cite{wiesehan2021negligible,imperatore2021reproducibility}, and the challenging separation of polaritonic and non-polaritonic effects in optical cavity setups \cite{schwennicke2024molecular,thomas2024non,michon2024impact}. Fundamentally, the mechanism behind the multitude of vibrational strong-coupling experiments remains thus unclear, despite the vast amount of theoretical developments in recent years \cite{campos2023swinging,fregoni2022theoretical,wang2021roadmap}.  

Typical Fabry-Pérot (FP) cavity setups comprise a large number of electrons and nuclei which couple to cavity photons. The enormous dimensionality of such cavity-molecule systems makes simulations, which treat electrons, nuclei and photons on the same quantum-mechanical footing, computationally extremely demanding. Hence, a multitude of approximate methods for solving the time-(in)dependent Schrödinger equation of the coupled electron-nuclear-photonic system have been developed \cite{flick2017atoms,fregoni2022theoretical,ruggenthaler2023understanding}. As of yet, however, there is no consensus in the theory community on parameter regimes (characterized by, e.g., coupling strength, number of molecules and cavity frequency) in which these approximations hold and where they break down. 

Specifically, two open questions are heavily debated in the current literature \cite{sidler2024unraveling,horak2025analytic,schäfer2020relevance,fiechter2024understanding,martinez2018can,galego2015cavity}:
When does the so-called dipole self-energy (DSE, or self-polarization term) become relevant such that it cannot be omitted in simulations? And: Can coupling to the cavity significantly alter the molecular electronic ground state under typical experimental conditions? Note that the two questions are connected since the dipole-self energy term mediates intra- and intermolecular electron-electron and electron-nuclear interactions which can impact the molecular electronic structure. While the first question addresses approximations to the light-mater interaction Hamiltonian, the second question relates to the appropriate approximations to the molecular wavefunction under strong coupling. In the following, we will thus briefly review both types of common approximations.

\begin{figure}
    \centering
    \includegraphics[width=\linewidth]{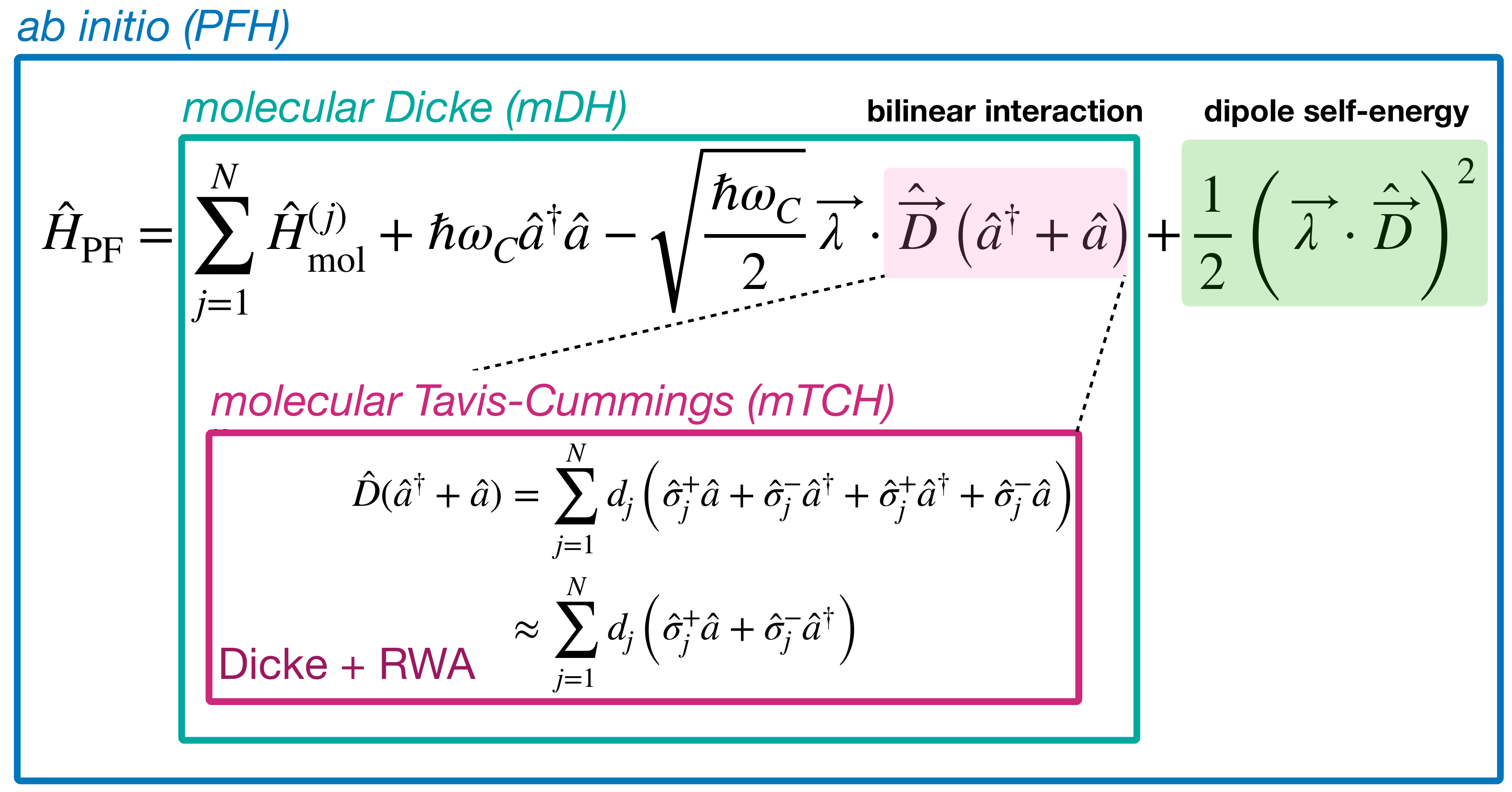}
    \caption{Hierarchy of approximations on the light-matter Hamiltonian level. The cavity mode displacement and momentum operator in \eqref{eq:full-ham} are expressed in ladder operators via $\hat q_C=i\sqrt{\hbar\omega_C/2}(\hat{a}^{\dagger}-\hat{a})$ and $\hat p_C=\sqrt{\hbar/2\omega_C}(\hat{a}^{\dagger}+\hat{a})$. $d_j$ denotes the magnitude of the transition dipole moment (associated with the excitation $\hat\sigma_j$) of the $j$-th molecule.}
    \label{fig:Figure-approx}
\end{figure}

It is instructive to categorize the approximations on the level of the light-matter Hamiltonian into a hierarchy which is displayed in Figure~\ref{fig:Figure-approx}. Starting point is the so-called Pauli-Fierz Hamiltonian, i.e., the minimal coupling Hamiltonian in Power-Zienau-Woolley (PZW) gauge which correctly describes the interaction between molecules and a quantized transverse electromagnetic field in the non-relativistic limit. \emph{Ab initio} treatments \cite{ruggenthaler2023understanding} of a FP cavity-molecule systems are usually based on the Pauli-Fierz Hamiltonian (PFH) with a single ``effective'' quantized mode of light,
\begin{gather}
    \hat H_{\mathrm{PF}}= \sum_{j=1}^N \hat H_{\mathrm{mol}}^{(j)} + \frac{\hat{p}_C^2}{2} +\frac{\omega_C^2}{2}\hat{q}_C^2-\omega_C \hat q_C\vec\lambda\cdot\hat{\vec D} + \frac{1}{2}\left(\vec\lambda\cdot\hat{\vec D}\right)^2 , \label{eq:full-ham}
\end{gather}
where $\hat q_C$ and $\hat p_C$ denote displacement coordinate and conjugate momentum of the cavity mode with frequency $\omega_C$ respectively, and $\vec\lambda$ is the light-matter coupling strength. Here and throughout the manuscript, atomic units are used if not mentioned otherwise.
Although the PFH describes exactly the interaction of matter and transverse electromagnetic modes in the absence of relativistic effects, and therefore it is termed “ab initio” in this sense, in most instances one does not consider the physical specifics of the cavity when determining the coupling strength $\vec\lambda$. Hence, in fact most applications of the PFH consider effective electromagnetic modes and treat $\vec\lambda$ as a model parameter (see, e.g., \cite{sidler2024unraveling,schnappinger2024molecular,castagnola2025realistic}).

In this form of the light-matter Hamiltonian, coupling between the molecular ensemble to the cavity field results in two distinct interaction terms: a bilinear interaction between cavity displacement and ensemble dipole $\hat{\vec D}$, $\hat{H}_{\mathrm{cav-mol}} \equiv -\omega_C \hat q_C \vec\lambda \cdot \hat{\vec D}$, and a quadratic term, mediating interactions between all molecular dipoles with themselves. The latter is known as the dipole self-energy (DSE) term, $\hat{H}_{\mathrm{dse}} \equiv \frac{1}{2}(\vec\lambda \cdot \hat{\vec D})^2$.
Intuitively, the DSE is an energetic penalty on the build-up of a macroscopic polarization within the cavity-molecule system. It thereby prevents molecules to polarize indefinitely inside the cavity 
\cite{schäfer2020relevance,rokaj2018light}.

When omitting the dipole self-energy term, one arrives at the molecular Dicke-Hamiltonian (mDH), which still captures exactly the bilinear interaction between the dipole moment of the molecular ensemble and the cavity mode. By further expressing the matter dipole moment in creation and annihilation operators and neglecting the counter-rotating terms, i.e., performing a rotating wave approximation (RWA) on the mDH, the molecular Tavis-Cummings Hamiltonian (mTCH) is obtained. The latter is often used when simulating many molecules in the ESC regime, e.g. in the context of exciton-polariton dynamics \cite{fregoni2022theoretical}.

We stress that these approximations are made solely on the level of the light-matter interaction Hamiltonian, and do not affect the molecular Hamiltonian $\hat H_{\mathrm{mol}}^{(j)}$ in \eqref{eq:full-ham}. In our definition of the Dicke and Tavis-Cummings Hamiltonian, the full complexity of $\hat H_{\mathrm{mol}}^{(j)}$ is retained. In the physics community, on the other hand, Dicke and Tavis-Cummings Hamiltonian usually refer to model systems in which the molecular Hamiltonian $\hat H_{\mathrm{mol}}^{(j)}$ is reduced to a two- or few-level system. To distinguish the approximations introduced in Figure~\ref{fig:Figure-approx} from these established quantum optics models, we use the terms \emph{molecular} Dicke and \emph{molecular} Tavis-Cummings Hamiltonian.

Understanding how an empty cavity mode, i.e., with zero photons present, affects molecules in their electronic ground state, is central to settling the mechanistic nature of VSC-modified reactivity \cite{campos2023swinging}. From the perspective of the light-matter Hamiltonian, modifications to the electronic ground state can either occur through the DSE, or through the off-resonant interactions induced by the bilinear term \cite{fischer2023beyond,galego2015cavity}. While both are included in the \emph{ab initio} Hamiltonian, the mDH neglects the impact of the DSE. Since the mTCH lacks both terms, it is not suitable for the description of molecular ground states under VSC, and is not discussed here.
 Nevertheless, one should remember that cavity reactivity in the single-molecule and collective VSC regime can also be modified by resonant effects, i.e. the presence of cavity modes resonant with specific vibrational modes of the reactive molecules \cite{campos2023swinging,sun2022suppression,li2021collective,sun2023modification,lindoy2023quantum}, which are accounted for by the bilinear coupling term of the mTCH. In this work, however, we aim to address, \emph{from first principles}, the nature and extent of modifications to the structure of the ground electronic state and to single-molecule properties under VSC across various levels of controlled approximation. Importantly, such modifications are independent of whether vibrational resonances are present \cite{galego2019cavity}.

After the choice of light-matter Hamiltonian, solving the Schrödinger equation for the cavity-molecule system usually starts by solving a time-independent Schrödinger equation for the electronic subsystem. In general, the molecular electronic wavefunction under strong coupling can be expanded in a set of eigenstates of the field-free electronic Hamiltonian. In the limit of a complete electronic basis, this approach is exact \cite{fiechter2024understanding}. In practice, however, a truncated basis is employed which limits this method's capability of capturing polarization effects of the cavity on electronic structure. If molecular electronic states are substantially modified by coupling to the cavity mode, the electronic wavefunction can no longer be represented in a truncated basis of field-free electronic states. This motivates a second family of computational approaches in which a modified electronic problem is solved that includes the interaction with the cavity and the DSE. In this modified electronic problem the cavity displacement is either treated as a dynamical variable (``QED'' methods, e.g., QED-CC, QED-ADC, QEDFT) \cite{haugland2020coupled,bauer2023perturbation,ruggenthaler2014quantum} or as a parameter (``cavity Born-Oppenheimer'' (cBOA) methods, e.g., cBOA-HF) \cite{flick2017atoms,schnappinger2023cavity}.

Finally, for simulations with more than one molecule the question of a suitable ansatz for the ensemble wavefunction arises since the size of the Hilbert space increases exponentially with the number of molecules. Within tree-tensor network methods such as multilayer multiconfiguration time-dependent Hartree (ML-MCTDH)\cite{wang2003multilayer,manthe2008multilayer,vendrell2011multilayer} this scaling is mitigated by employing variationally-optimal tensor decomposition, allowing for accurate, numerically exact treatment of intermolecular correlations. Alternatively, recent publications \cite{schnappinger2023cavity,sidler2024unraveling,sidler2024collectively} have advocated the use of a mean-field ansatz which is computationally much cheaper but neglects such correlations. 

Overall, when simulating a concrete cavity-molecule system under VSC, one faces a number of ambiguities regarding approximations to the Hamiltonian and wavefunctions. On the level of the Hamiltonian, it is still uncertain whether the use of the full PFH is required, or a mDH suffices. On the level of the wavefunction, it is unsure whether employing electronic field-free states
(their potential energy surfaces and non-adiabatic couplings) gives an accurate description, or if using cavity-modified electronic states is more appropriate. Lastly, for ensembles, it is unclear to what extent correlation should be included in the
wavefunction ansatz, both at the light-matter and intermolecular levels.
Identifying the relevant parameter regimes, in which these approaches are accurate, requires simulations that rely on minimal approximations, and benchmark studies comparing different theoretical methods across a range of coupling strengths, molecular concentrations, and cavity frequencies.

Here, simulations of cavity-ensemble systems which treat photons, nuclei and electrons on equal, quantum-mechanical footing are performed to assess the relevance of the dipole-self energy term, cavity-induced correlations and modifications to molecular structure. First, a physical model is devised which is complex enough to capture the rich interplay between molecular and photonic degrees of freedom (DOFs), yet simple enough to afford simultaneous quantum-mechanical treatment of all DOFs. To this end, we consider an ensemble of Shin-Metiu (SM) systems \cite{shin1995nonadiabatic} coupled to a single effective quantized cavity mode described by the PFH in length gauge and in dipole approximation \cite{ruggenthaler2023understanding}. Previous studies based on a similar model could gain valuable insight into polaritonic chemistry \cite{galego2019cavity,sidler2024unraveling,li2021cavity}. Besides, the single electron of an individual SM system can be easily treated alongside nuclear DOFs within variationally optimal wavepacket propagation methods such as multilayer multiconfiguration time-dependent Hartree (ML-MCTDH) due to absence of fermionic antisymmetry. 
This allows us to retain an exact description of the molecular SM Hamiltonian in \eqref{eq:full-ham}, and cleanly compare the \emph{ab initio} PFH with the mDH in which the DSE has been omitted. We consider both fixed-orientation and freely rotating SM systems to investigate the role of molecular rotations. This becomes particularly important with the advent of gas-phase strong-coupling experiments \cite{wright2023rovibrational}.

We employ the ML-MCTDH algorithm \cite{wang2003multilayer,manthe2008multilayer,vendrell2011multilayer} to compute the correlated
ground state of the coupled rotational-vibrational-electronic-photonic ensemble under VSC. Comparison with simulations based on approximate Hamiltonians and wavefunctions sheds light on the role of cavity-induced polarization, DSE and intermolecular correlation over a large parameter space spanning weak to ultrastrong coupling, from the single-, few-molecule and up to the collective regime.

In a single and few-molecule coupling scenario, we find pronounced reorientation and small distortions along vibrational coordinates at the onset of the ultrastrong coupling (USC) regime, while electron densities do not show significant alterations up to very large per-molecule coupling strengths. In the USC regime, we can also confirm the importance of including the DSE to ensure stability of the ground state. Crucially, our numerically exact simulations show that the response of molecular DOFs and the onset of instability due to omission of the DSE, depend on the per-molecule, not the collective coupling strength.  

While single- or few-molecule USC using FP cavities is a hypothetical scenario, the collective strong and ultrastrong coupling regime are experimentally accessible with macroscopic numbers of molecules present in the FP cavity volume. The scaling of local, i.e., intramolecular, cavity-induced modifications with the per-molecule coupling strength implies that they do not profit from collective enhancement. Accordingly, reorientation, and nuclear and electronic polarization effects decay rapidly with increasing number of molecules for a fixed collective coupling strength. Thus, numerically exact simulations of the \emph{ab initio} Hamiltonian corroborate the negligible impact of the cavity on the local molecular ground states in the thermodynamic limit ($N\to\infty$). Moreover, simulations with the mDH converge to the same result.
Nonetheless, we highlight the importance of wavefunction correlation among molecules and between cavity and molecules for accurately describing global properties of the coupled light-matter system.
\section{Theory}
 We consider a gas of $N$ SM systems with electronic, nuclear and rotational DOFs coupled to a single-mode FP cavity, either described by the PFH or mDH, as introduced around \eqref{eq:full-ham} (also see Figure~\ref{fig:Figure-approx}).
The molecular Hamiltonian $\hat H_{\mathrm{mol}}^{(j)}$ of an individual SM system reads
\begin{eqnarray}
    \hat H_{\mathrm{mol}}^{(j)} = &&\frac{\hat{J}_j^2}{2I} -\frac{\nabla_{r_j}^2}{2} -\frac{\nabla_{R_j}^2}{2M} + \hat V_{\mathrm{sm},j} \label{eq:sm-ham}
\end{eqnarray}
with 
\begin{eqnarray}
    \hat V_{\mathrm{sm},j} = &&\frac{1}{|\hat{R}_j-L/2|} + \frac{1}{|\hat{R}_j+L/2|} -\frac{\mathrm{erf}(|\hat{R}_j-\hat{r}_j|/R_l)}{|\hat{R}_j-\hat{r}_j|} \nonumber\\&&- \frac{\mathrm{erf}(|\hat{r}_j-L/2|/R_f)}{|\hat{r}_j-L/2|} - \frac{\mathrm{erf}(|\hat{r}_j+L/2|/R_r)}{|\hat{r}_j+L/2|}.
\end{eqnarray}
Each SM model system \cite{shin1995nonadiabatic} (Figure~\ref{fig:Figure1}a) consists of an electron with position operator $\hat r_j$, a mobile nucleus with position operator $\hat R_j$, and two fixed nuclei at position $-L/2$ and $L/2$. The mobile nucleus interacts with the fixed nuclei through regular Coulomb potentials, whereas all attractive interactions between nuclei and electrons are given by ``softened'' Coulomb potentials which are parametrized by $R_f,R_m$ and $R_r$. Charge $Z=1$ and mass $M=1836$ are chosen for the mobile ion, the momentum of inertia $I$ of dihydrogen is used. The parameters of the SM model are taken from Ref.~ \cite{sidler2024unraveling}, and are summarized in Tab.~\ref{tab:model-params}.

\begin{table}
    \centering
    \begin{tabular}{cc}\toprule
        parameter & value \\ \midrule
        $L$ & 9.45~Bohr \\  
        $R_f$ & 1.511~Bohr   \\  
        $R_m$ & 1.511~Bohr  \\  
        $R_r$ & 1.511~Bohr \\  
        $I$  & $4.58 \times 10^{-48} \, \text{kg} \cdot \text{m}^2$\\  
        $M$ & 1836~amu \\ 
        $Z$ & 1 \\
        $\omega_C$ & 6.27 mH $\approx$ 1380\,$\mathrm{cm}^{-1}$\\
        $\lambda$ & 0 -- 0.045  \\
        $\epsilon_C$ &  0 -- 1.3\,V/nm \\
        \bottomrule
    \end{tabular}
    \caption{Cavity ($\lambda$, $\omega_C$, $\epsilon_C$) and SM model parameters ($L$, $R_f$, $R_m$, $R_r$, $I$, $M$, $Z$), which were adopted from Ref.~ \cite{sidler2024unraveling}. All quantities are defined in the text.}
    \label{tab:model-params}
\end{table}

Note that nuclei and electron of the SM model are restricted to move in one dimension only. We stay consistent with this underlying assumption by neglecting coupling between the rotational ($\hat \theta_j$) and electronic motion ($\hat r_j$), such that all particles of the SM model stay on a line during rotation. In addition to that, the small centrifugal couplings between rotation and vibration are neglected as well in the kinetic energy operator of the model.

To couple the ensemble of SM molecules with the cavity mode, we write the ensemble dipole moment as 
$\hat{\vec{D}}=\sum_{j=1}^N \hat \mu_{\mathrm{bf}}^{(j)}\cdot \vec{u}_j(\theta_j)$, where $\vec{u}_m(\theta_j) = \left(\cos(\theta_j),\,\sin(\theta_j)\right)^T$ describes the orientation of the individual molecular dipoles with respect to the cavity polarization direction (see Figure~\ref{fig:Figure1}b), and $\hat\mu_{\mathrm{bf}}^{(j)}$ is the body-fixed dipole operator of the $j$-th SM molecule. Thus, the dot product $\vec\lambda\cdot\hat{\vec D}$ which appears in the bilinear term and quadratic self energy term is given by
\begin{gather}
    \vec\lambda\cdot\hat{\vec D} = \lambda\sum_{j=1}^N \cos(\theta_j)\left(\hat{R}_j - \hat{r}_j\right),
\end{gather}
with the light-matter coupling strength $\vec\lambda = \vec e_C \lambda$, where $\vec e_C$ denotes the cavity mode polarization direction. $\lambda$ can be expressed in terms of the (effective) cavity mode volume $V$ through $\lambda=1/\sqrt{\varepsilon_0 V}$. The corresponding vacuum electric field strength is given by $\epsilon_C=\sqrt{\omega_C/2}\lambda$.

Furthermore, the exact treatment of the dipole self energy term, i.e., the squared expression $(\vec\lambda\cdot\hat{\vec D})^2$, is also straightforward since it corresponds to a sum of products of electronic, nuclear and rotational coordinates,
\begin{eqnarray}
    \left(\vec\lambda\cdot\hat{\vec D}\right)^2 =  &&\lambda^2\sum_{j=1}^N \cos^2(\theta_j)( \hat{R}_j^2 - 2 \hat{R}_j \hat{r}_j + \hat{r}_j^2)\nonumber\\
    &&+ 2\lambda^2\sum_{j=1}^N\sum_{i<j}^N \cos(\theta_i)\cos(\theta_j)\times \nonumber\\&&\left(\hat{R}_i\hat{R}_j + \hat{r}_i \hat{r}_j - \hat{R}_i \hat{r}_j - \hat{R}_j \hat{r}_i\right), 
\end{eqnarray}
which is easily handled by our ML-MCTDH implementation \cite{beck2000multiconfiguration,vendrell2011multilayer}.

\section{Methodology}
\begin{figure}
    \centering
    \includegraphics[width=\linewidth]{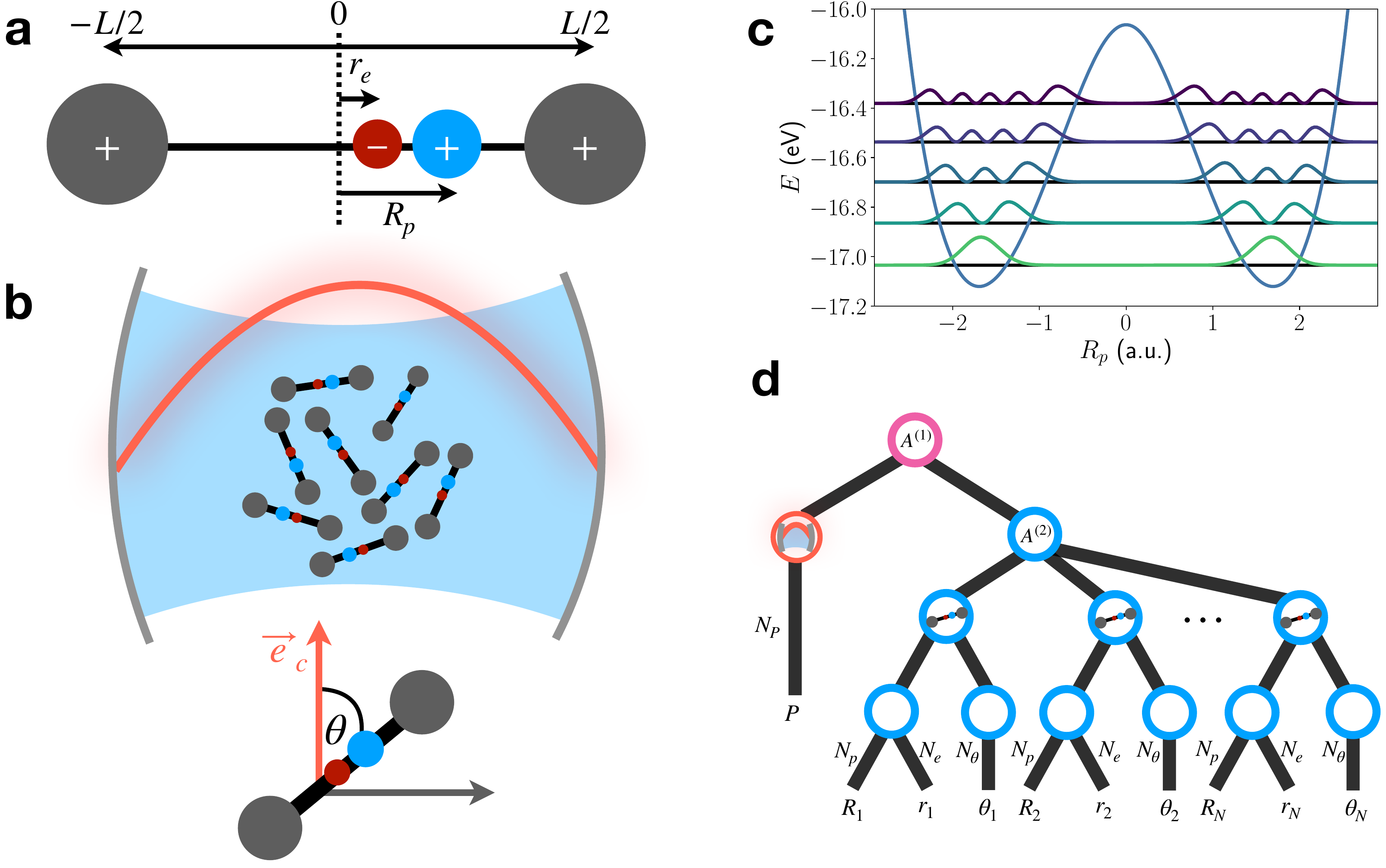}
    \caption{Schematic of cavity-coupled Shin-Metiu (SM) ensemble and computational strategy. (a) Each SM ``molecule'' consists of two fixed positive charges at $-L/2$ and $L/2$ and a mobile electron and proton with positions $r_e$ and $R_p$, respectively. (b) An ensemble of $N$ gas-phase SM model systems couples collectively to an infrared mode of a Fabry-Pérot microcavity. The molecules can rotate freely, characterized by the polar angle $\theta$ which is measured with respect to the cavity-mode polarization axis $\vec e_C$. (c) Ground-state potential energy curve of the bare SM model system superimposed with the first five vibrational levels and corresponding nuclear densities. The cavity is tuned resonant to the fundamental vibrational transition. (d) The polaritonic ground state is computed by imaginary time propagation of the full-dimensional (photonic-electronic-rovibrational) wavefunction in tree-tensor network format. An exemplary tree is shown in ML-MCTDH tree notation.}
    \label{fig:Figure1}
\end{figure}

To meet the computational challenge of treating the high-dimensional cavity-ensemble wavefunction $\psi(q_C,r_1,\dots,r_N,R_1,\dots,R_N,\theta_1,\dots,\theta_N)$ we employ the ML-MCTDH method which represents the wavefunction in a compact tree-tensor network (or hierarchical Tucker) format \cite{wang2003multilayer,manthe2008multilayer,vendrell2011multilayer}. This wavefunction ansatz can be graphically represented in a tree diagram where each node corresponds to a coefficient tensor in the tensor-network wavefunction expansion (Figure~\ref{fig:Figure1}d, more details in Supporting Information). The use of such tensor network-type wavefunctions becomes mandatory for full quantum-mechanical treatment of the present cavity-ensemble problem: representing the cavity-ensemble wavefunction on a direct product grid of basis sets for photonic, electronic, vibrational and rotational DOFs would require $N_P\cdot (N_p\cdot N_e\cdot N_{\theta})^N$ coefficients where $N_P$, $N_e$, $N_p$, $N_{\theta}$ denote the size of these basis sets (respectively). For $N=10$ and the grids used in the present calculations, approximately $3.8\times10^{30}$ coefficients would be required. ML-MCTDH mitigates this unfavorable scaling through variationally optimal tensor decomposition, thus allowing to study larger molecular ensembles coupled to a cavity, even in the strong to ultrastrong coupling regime, while accurately capturing correlation between all constituents of the system \cite{vendrell2018coherent}.

Besides considering coupled SM sub-systems with full rotational, vibrational and electronic freedom, no further approximations are made: our approach does not rely on a cavity Born-Oppenheimer ansatz, nor any other adiabatic separation between slow and fast DOFs, and it does not require pre-computed dipole moments and potential energy surfaces. Instead, all DOFs are treated on equal footing by representing kinetic energy operators, potentials and dipole operators directly on (real-space) grids. Moreover, the flexibility and compactness of the tree tensor-network wavefunction ansatz enables us to converge to the fully correlated cavity-ensemble wavefunction. This goes well beyond previous resorts to single-Hartree product and mean-field approximations \cite{schnappinger2023cavity,sidler2024unraveling,horak2025analytic}. Overall, our methodology provides access to
\textit{first principles} simulations of molecules coupled to a cavity-mode, i.e., treatment of the single-mode PFH and all its constituents --photons, nuclei and electrons-- entirely based on quantum mechanics \cite{ruggenthaler2023understanding}.

With this computational toolbox at hand, we focus here on polaritonic ground-state properties and their dependence on coupling strength and ensemble size, comparing numerically exact and approximate calculations. The polaritonic ground state is obtained by propagating the imaginary-time Schrödinger equation
$\frac{\partial }{\partial t}|\Psi\rangle = -\hat H |\Psi\rangle$ with the Heidelberg ML-MCTDH package starting from a Hartree product wavefunction, i.e., by performing a wavepacket relaxation \cite{beck2000multiconfiguration}. An energy difference $\Delta E \leq 10^{-8}$\,au between consecutive steps of the relaxation was used as the convergence criterion. For more computational details we refer the reader to the Supporting Information.

 The bare SM system possesses a symmetric double well potential in the electronic ground state (Figure~\ref{fig:Figure1}c), modeling an isomerization or proton-coupled electron-transfer reaction. This potential energy curve supports pairs of nearly-degenerate vibrational levels which are split by a very small tunneling splitting. Here, the cavity frequency is tuned resonant to the transition between the first and second pair of nearly-degenerate vibrational states at $\omega_{\mathrm{v}}=1380$\,cm$^{-1}$. This value corresponds to a typical vibrational mode frequency in the IR range. Note, however, that modifications of the ground state under strong coupling are not tied to a resonance between cavity and molecular transition, as discussed above.
 
Changes to the ground-state properties are monitored by calculating expectation values $\langle \psi|\hat{\mathcal{O}}|\psi\rangle$ for the converged ground-state wavefunction $|\psi\rangle$. We study here ground-state modifications on the level of individual molecules in the ensembles, i.e., $\hat{\mathcal{O}}\equiv \hat{\mathcal{O}}_j$ depends only on one ore more DOFs of a single molecule with index $j$. Since  molecules are identical in our model, ensemble averages $\sum_{j=1}^N \langle \psi|\hat{\mathcal{O}}_j|\psi\rangle/N$ and expectation values of a single representative molecule $\langle\psi|\hat{\mathcal{O}}_j|\psi\rangle$ coincide. This has been checked in the SI (see Figure~S2). Operators acting on rotational coordinates ($\mathrm{cos}^2(\theta)$), electronic coordinates ($\hat r_e^2$), vibrational coordinates ($\hat R_p^2$) and both electronic-vibrational coordinates ($\hat \mu^2=(\hat R_p-\hat r_e)^2$) are employed to quantify these changes. The ground-state expectation value of linear functions of $\hat r_e$, $\hat R_p$ and $\hat q_C$ vanishes due to inversion symmetry of the cavity SM Hamiltonian (cf. equations \eqref{eq:full-ham} and \eqref{eq:sm-ham}). For the same reason, the total dipole moment of the SM system  studied here is zero. Accordingly, $\langle \hat \mu^2\rangle$ and $\langle \hat q_C^2\rangle$ correspond to the ground-state fluctuations of the dipole and displacement field.

\section{Results}
\subsection{Cavity-induced changes of molecular structure}\label{sec:changes}
The effect of the cavity on electronic, nuclear and rotational DOFs in a gas of SM systems described by the full Hamiltonian \eqref{eq:full-ham} is studied as a function of the coupling strength $\lambda$. In typical vibrational strong coupling (VSC) experiments with FP cavities, $\lambda$ can be estimated from the cavity mode volume $V$ as $\lambda = 1/\sqrt{\epsilon_0 V}$, which must be at least on the order of $\lambda_C^3$, where $\lambda_C$ is the cavity wavelength  \cite{li2021collective}. For an IR cavity with $\lambda_C = 1~\mu$m, this yields $\lambda \sim 10^{-8}$.

Strong coupling is identified spectroscopically via the Rabi splitting, scaling as $\sqrt{N}$. Achieving experimental splittings with such small per-molecule coupling would require simulating $10^6$–$10^{12}$ molecules, which is computationally infeasible. Therefore, most theoretical studies use an effective coupling strength
$\lambda_{\mathrm{eff}} = g / (\mu_{01} \sqrt{N} \sqrt{\omega_v/2})$, where $g$ is the collective coupling, $\mu_{01}$ the transition dipole moment, and $\omega_v$ corresponds to the frequency of the molecular transition coupled to the cavity. This allows reproducing a desired Rabi splitting $\hbar\Omega_R = 2g$ with a smaller $N$, though at the cost of artificially increasing the per-molecule coupling strength. Here, $N$ becomes a convergence parameter that should be increased to approach the thermodynamic limit \cite{ribeiro2022multimode,sokolovskii2024one}.
To match VSC conditions, typical values are $\lambda_{\mathrm{eff}} \sim 10^{-4}$–$10^{-3}/\sqrt{N}$. In the following, we also explore stronger coupling regimes, including the ultrastrong coupling regime, defined by $g/\omega_v \geq 0.1$  \cite{hoblos2025does}, corresponding to $\lambda_{\mathrm{eff}} \gtrsim 0.01/\sqrt{N}$ in our model. Equivalently, in an IR cavity with $\omega_C=1380$\,cm$^{-1}$, the onset of the single-molecule USC regime occurs at a vacuum cavity field strength of $\epsilon_C=0.29$\,V/nm, while the largest per-molecule coupling strength corresponds to $\epsilon_C=1.3$\,V/nm. These coupling strengths are out of reach in FP configurations with transverse light-matter coupling only \cite{de2025there,fregoni2022theoretical}, but are considered here as we investigate the theoretical limit of the PFH.

\begin{figure}
    \centering
    \includegraphics[width=\linewidth]{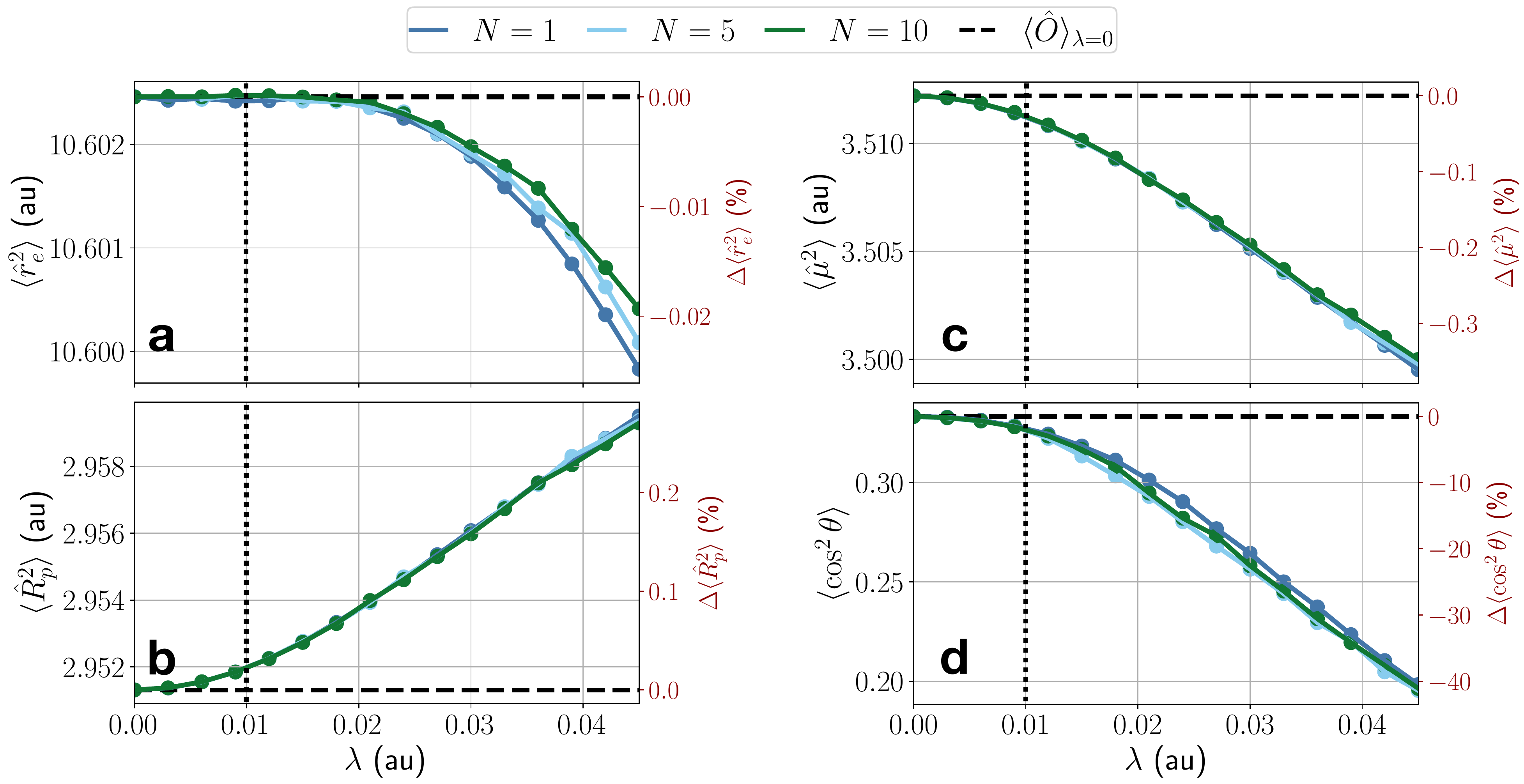}
    \caption{Cavity-induced modifications of local observables for increasing per-molecule coupling strengths ($\lambda$). Changes are given in absolute terms (left axis) and relative to the bare-molecule value, $\Delta \langle\mathcal{\hat O}\rangle = (\langle\mathcal{\hat O}\rangle_{\lambda} - \langle\mathcal{\hat O}\rangle_{\lambda=0})/\langle\mathcal{\hat O}\rangle_{\lambda=0}$ (right axis, in dark-red). The horizontal dashed line indicates the respective bare-molecule value $\langle\mathcal{\hat O}\rangle_{\lambda=0}$, the vertical dotted line marks the onset of the single-molecule USC regime. In this and the following Figures, symbols (here: circles) mark the computed data points, curves are included as a visual aid.}
    \label{fig:Figure2}
\end{figure}

Keeping this in mind, we now turn to Figure~\ref{fig:Figure2} in which no $1/\sqrt{N}$ scaling to the coupling strength has been applied, i.e., the per-molecule coupling strength $\lambda$ is independent of ensemble size $N$. Most interestingly, the modifications to molecular DOFs in the ground state under strong-light matter coupling are independent of the ensemble size in this case. Only a very weak dependence on $N$ is found in the electronic observables at very large coupling strengths. This indicates that ground-state properties depend on the \textit{per-molecule coupling} strength, $\lambda$, and not the \textit{collective} coupling strength, which scales with $\sqrt{N}$.

When increasing the coupling strength $\lambda$ of each individual molecule in the ensemble toward (artificially) high values, all DOFs display cavity-induced changes with respect to the uncoupled molecule case. Yet, the degree of these changes depends strongly on the type of DOF. Molecular rotations exhibit the most pronounced modifications in Figure~\ref{fig:Figure2}d: increasing the coupling strength, changes the alignment $\langle\mathrm{cos}^2(\theta)\rangle$ with the cavity polarization axis from the isotropic value of $1/3$ in the uncoupled case to approximately 0.20 for $\lambda=0.045$. Vibrational DOFs display the second-strongest modifications, albeit at a much smaller relative scale (see percentage axis on the right in Figure~\ref{fig:Figure2}b). At the largest coupling strength $\langle R_p^2\rangle$ is about 0.3\% larger than its value for the bare molecule (i.e., at $\lambda=0$). Below the onset of the single-molecule USC regime, these deviations are even smaller, below 0.05\%. The response of electronic DOFs is very small in absolute and in relative terms over a large range of coupling strengths (up to $\lambda\approx 0.02$), and even for the largest coupling strengths $\langle \hat r_e^2\rangle$ deviates by less than 0.02\% from its value at $\lambda=0$ (Figure~\ref{fig:Figure2}a). Moreover, the square of the dipole moment decreases with increasing coupling strengths, transitioning from a quadratic behavior at lower coupling strengths to a linear decrease in the USC (Figure~\ref{fig:Figure2}c). In total, changes in the squared dipole moment are on the order of 0.1\% of the bare-molecule value. Note also that increasing the number of molecules $N$ reduces the per-molecule change of polarizability (as reflected in $\langle\mu^2\rangle$) and electronic and nuclear position variances in Figures~\ref{fig:Figure2} and S3.

When molecular orientations are fixed to be aligned with the cavity mode polarization direction, the cavity-induced modifications to molecular structure are slightly more pronounced (cf.~Figure~S3), as the interaction between molecules and cavity is maximized. Yet, the relative response of electronic DOFs is still 1-2 orders of magnitude smaller than the response of vibrational DOFs, and molecular structure-modifications scale, as in the case of freely rotating molecules, with the per-molecule coupling strength until the onset of the USC regime.

\begin{figure}
    \centering
    \includegraphics[width=\linewidth]{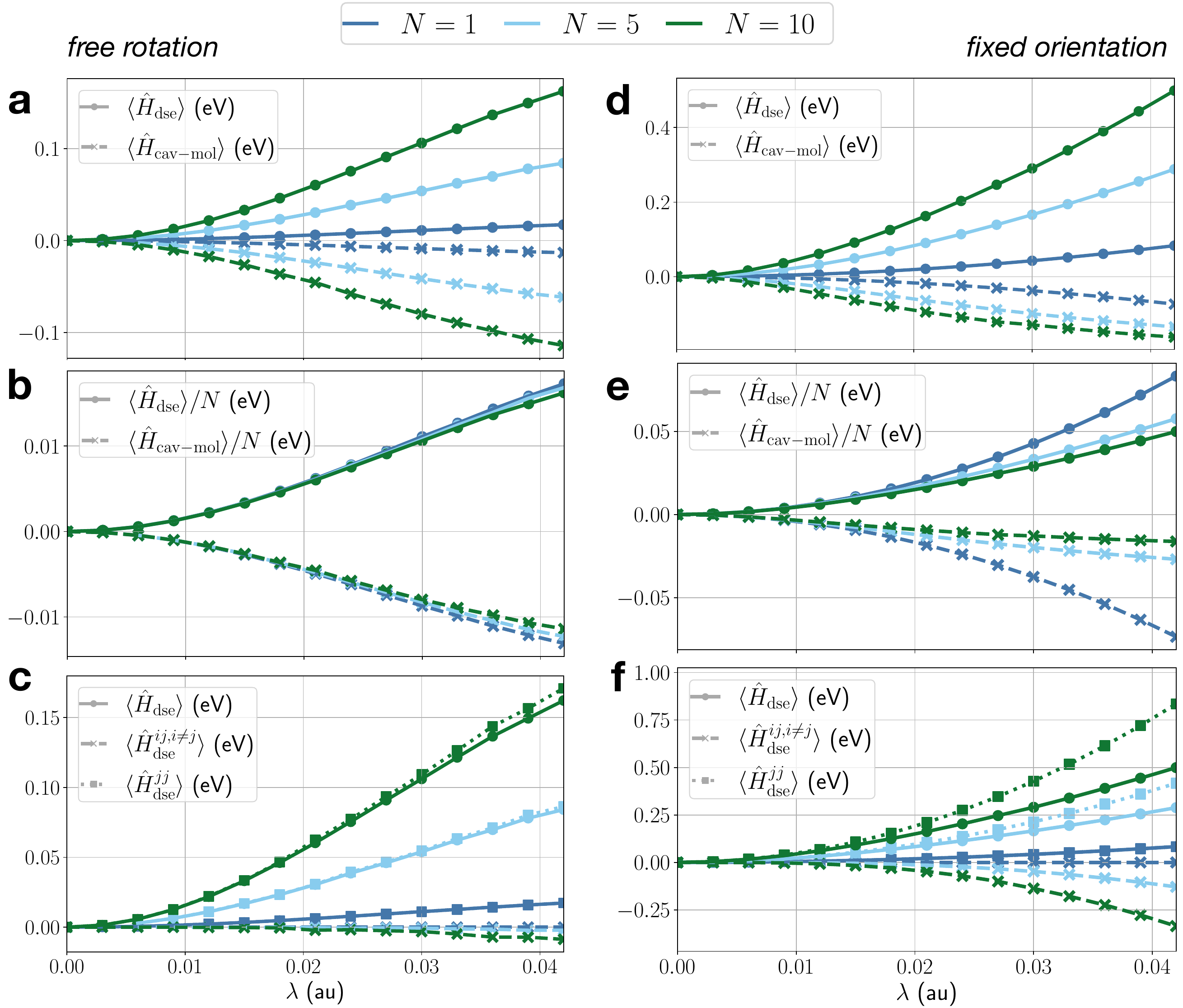}
    \caption{Contributions of dipole self-energy and bilinear term to ground state energy. (a) Total contributions of dipole self-energy $\langle \hat H_{\mathrm{dse}}\rangle$ and bilinear term $\langle \hat H_{\mathrm{cav\text{-}mol}} \rangle$ for $N=1,5,10$ as a function of per-molecule coupling strength $\lambda$. (b) Same as (a) but weighted by number of molecules. (c) Single-body $\langle\hat H_{\mathrm{dse}}^{jj}\rangle$ and two-body contributions  $\langle\hat H_{\mathrm{dse}}^{ij,i\neq j}\rangle$ to dipole self-energy as a function of per-molecule coupling strength $\lambda$. (d)-(f) Same as (a)-(c) but molecular orientations are fixed and aligned with cavity mode polarization axis.}
    \label{fig:Figure3}
\end{figure}

So far we have seen that there are no noticeable modifications of intramolecular ground-state properties, unless \emph{each individual molecule }in the ensemble is ultrastrongly coupled to the cavity. In this regime, molecular DOFs respond to reduce the polarizability in the ground state, which has already been observed in previous theoretical studies \cite{schnappinger2024molecular,philbin2023molecular}. This is reflected in the reduction of the body-fixed squared-dipole and the rotation out of the cavity polarization direction with increasing coupling strength. The latter reduces the interaction between the molecular ensemble and the cavity field altogether. This ground-state response of cavity-molecule systems can be explained by the increasingly dominant energy contribution from the DSE at such extreme coupling strengths. As shown in Figure~\ref{fig:Figure3}a and S4, the DSE contribution $\langle \hat H_{\mathrm{dse}}\rangle$ to the total ground-state energy exceeds the contribution of the bilinear cavity-molecule interaction $\langle \hat H_{\mathrm{cav-mol}}\rangle$. The latter term stabilizes the cavity-molecule system with increasing molecular polarization, whereas the DSE penalizes it. By lowering $\langle \mu^2\rangle$ and the alignment with the cavity polarization axis $\langle\mathrm{cos}^2(\theta)\rangle$, the energy penalty resulting from  $\hat H_{\mathrm{dse}}$ is thus minimized.

While both $\langle \hat H_{\mathrm{dse}} \rangle$ and $\langle \hat H_{\mathrm{cav\text{-}mol}} \rangle$ increase with the number of molecules (assuming no $1/\sqrt{N}$ scaling of the coupling strength), the per-molecule contribution $\langle \hat H_{\mathrm{dse}} \rangle/N$ remains essentially independent of $N$ across the entire range of examined coupling strengths, when molecules can freely rotate (Figure~\ref{fig:Figure3}a-b). In the case of fixed orientations (e.g. in condensed phase or a crystal) where all molecular dipoles are aligned with the cavity mode polarization axis, the same is observed up to a very large per-molecule coupling strength of $\lambda \approx 0.015$ (Figure~\ref{fig:Figure3}d-e).

Hence, the bilinear term energy contribution and DSE both behave as \emph{extensive} quantities until the onset of the single-molecule USC regime. Therefore, the \emph{local} (i.e., per-molecule) energetic changes do not display collective enhancement, except if \emph{each molecule} is already in the USC regime. In particular, this will also prohibit noticeable local modifications in the \emph{collective} USC, as we will discuss in Section~\ref{sec:largeN}. 

This observation demands closer analysis. To that end, we decompose the dipole self-energy (DSE) term into one- and two-molecule contributions, expressed as $\langle \hat H_{\mathrm{dse}} \rangle = \langle \hat H_{\mathrm{dse}}^{jj}\rangle+ \langle \hat H_{\mathrm{dse}}^{ij} \rangle$, where $\hat H_{\mathrm{dse}}^{ij} = \frac{\lambda^2}{2} \sum_{i \neq j} (\vec\lambda \hat{\vec{\mu}}_i)( \vec\lambda \hat{\vec{\mu}}_j)$ captures two-body interactions, and $\hat H_{\mathrm{dse}}^{jj} = \frac{\lambda^2}{2} \sum_{j} (\vec\lambda \hat{\vec{\mu}}_j)^2$ represents one-body terms.
Beyond a coupling strength of $\lambda\approx0.015$ (cf. Figure~\ref{fig:Figure3}f), the two-body contribution increases substantially and competes with the one-body terms of the DSE. These terms have opposite sign, resulting in a net increase of the DSE with $\lambda$. Consequently, up to $\lambda\approx0.015$ we find a one-body regime where uncorrelated one-body terms of the DSE and bilinear interaction dominate. This is reflected in the additivity of DSE and cavity-molecule interaction with respect to the number of molecules (Figure~\ref{fig:Figure3}d-e), and the scaling of local molecular modifications with the per-molecule coupling strength (cf.~Figure~S3). Importantly, both quantities increase also with $\lambda$ but with opposite signs. This competition between uncorrelated parts of DSE and the bilinear interaction determines the extent of local molecular modifications. Since the DSE remains slightly larger, the polaritonic ground state is mainly shaped by minimization of $(\vec\lambda \hat{\vec{\mu}}_j)^2$ of each individual molecule, forcing them to reduce polarizability along the cavity axis.

Beyond the onset of USC ($\lambda>0.015$), the cavity-ensemble system transitions into a many-body regime,  where intermolecular correlations either induced by two-body terms in DSE or indirectly via interaction of each molecule with the cavity DOF, become important. In this regime, the linear scaling  of $\langle \hat H_{\mathrm{dse}} \rangle$ and $\langle \hat H_{\mathrm{cav\text{-}mol}} \rangle$ with the number of molecules $N$ breaks down (cf.~Figs.~\ref{fig:Figure4}b and \ref{fig:Figure4}e). Still, local molecular modifications in Figure~S3 display ensemble size-independent behavior up to much larger coupling strengths ($\lambda\approx0.03$), indicating that intermolecular correlations play a minor role for intramolecular observables.

When the molecules can freely rotate, the per-molecule coupling strength is effectively reduced, at small couplings due to the isotropic averaging, and at larger coupling due to the pronounced reorientation of molecules away from the cavity mode polarization direction (cf.~Figure~\ref{fig:Figure2}d). Therefore,  $\langle \hat H_{\mathrm{dse}} \rangle$ and $\langle \hat H_{\mathrm{cav\text{-}mol}} \rangle$ are much smaller, approximately $1/3$ of the corresponding values obtained for fixed orientation. This prohibits the system from transitioning into a many-body regime as the necessary interaction strength with the cavity mode cannot be reached. Hence, the one-body part of the DSE dominates across the entire range of examined coupling strengths %
(cf.~Figure~\ref{fig:Figure3}c). Note that the rotational SM model has only one polarizable axis. Based on our results, real molecular systems in the gas phase will tend to expose their least polarizable axis to the cavity at high coupling strengths, in agreement with the results of \cite{philbin2023molecular}.

\subsection{Relevance of dipole self-energy}
Given the importance of the balance between DSE and bilinear interaction, we expect large deviations when omitting $\hat{H}_{\mathrm{dse}}$ in the coupling regime where ground-state modifications become noticeable (fractions of a percent). Indeed, we find pronounced differences in all observables in Figure~\ref{fig:Figure4}a-d between calculations with and without $\hat{H}_{\mathrm{dse}}$ starting from a per-molecule coupling strength $\lambda\approx 0.01$. Without rotation, deviations appear slightly earlier, at $\lambda\approx 0.006$ (see Figure~S5).

In the ground-state of the full PFH, the bilinear cavity-molecule interaction is fully compensated by the uncorrelated part of the DSE. The impact of this missing compensation in simulations without $\hat H_{\mathrm{dse}}$ is twofold. On the one hand, polarization of the molecules now has a stabilizing effect on the ground state. Correspondingly, increasing $\lambda$ results in an increase of $\langle \mu^2\rangle$ (Figure~\ref{fig:Figure4}c) and polarizable molecular axes orient toward the cavity polarization axis (Figure~\ref{fig:Figure4}d) to maximize their interaction with the cavity mode. On the other hand, local observables display an artificial collective scaling in the absence of the DSE, as seen in Figure~\ref{fig:Figure4}. As opposed to Figure~\ref{fig:Figure2} one finds a strong dependence of $\langle \hat{\mathcal{O}} \rangle_{\lambda}$ on the ensemble size $N$ (without employing a $1/\sqrt{N}$ scaling of the coupling strength), where cavity-induced modifications in the ground state are enhanced for larger $N$. {Without compensation by the predominantly uncorrelated $\langle \hat H_{\mathrm{dse}}\rangle$, $\langle \hat H_{\mathrm{cav-mol}}\rangle$ grows too quickly and promotes the build-up of correlation among molecules: the linear scaling of the bilinear interaction with $N$ breaks down even for freely rotating molecules from $\lambda\approx 0.01$ onward in Figure~\ref{fig:Figure4}e.

\begin{figure}
    \centering
    \includegraphics[width=\linewidth]{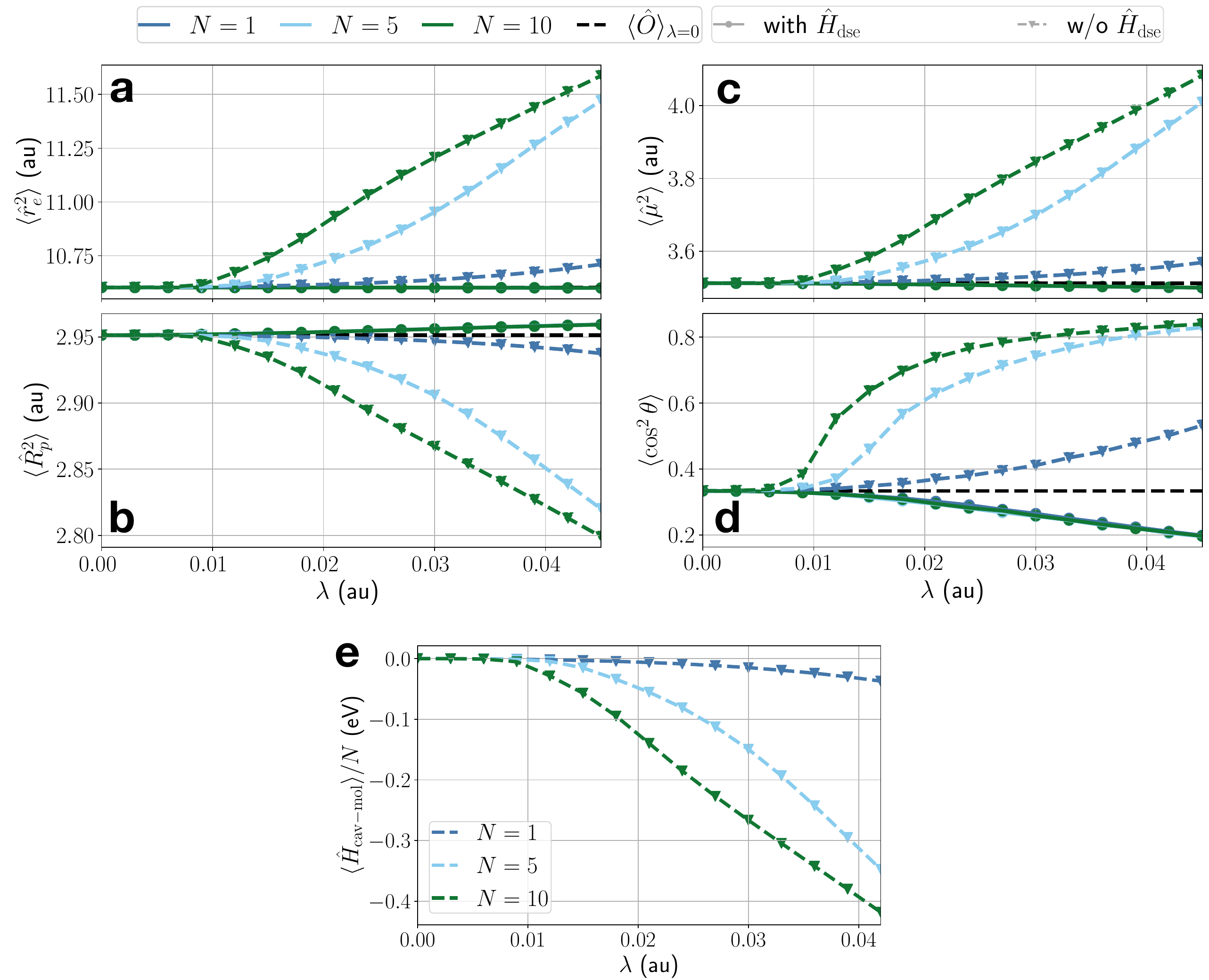}
    \caption{Impact of neglecting dipole-self energy terms for increasing coupling strength. (a)-(d) Modification of local observables $\langle\hat{\mathcal{O}}\rangle_{\lambda}$. The results with dipole-self energy from Figure~\ref{fig:Figure2} are shown in solid lines for comparison. (e) Single molecule contribution to bilinear term $\langle\hat{H}_{\mathrm{cav-mol}}\rangle$ for varying $N$.}
    \label{fig:Figure4}
\end{figure}

We stress that these findings only apply to the PFH when used to describe FP cavities at artificially high coupling strengths. Strong coupling to plasmonic nanocavities is qualitatively different, as the ``light''-molecule interaction is dominated by longitudinal Coulomb interactions and not by coupling to purely transverse modes. As pointed out by Fregoni~\textit{et al.} \cite{fregoni2022theoretical} and recently in Ref.~\cite{de2025there}, the Hamiltonian should therefore not include the DSE in the latter case. Moreover, the coupling strengths at which substantial differences appear in the reported simulations are significantly larger than feasible values in FP cavity setups \cite{de2025there}, making the single- and few-molecule ultrastrong regime with purely transverse modes a hypothetical scenario.

\subsection{Mean-field approximations to cavity-ensemble wavefunction}
\label{sec:resC}

\begin{figure*}
    \centering
    \includegraphics[width=1.0\linewidth]{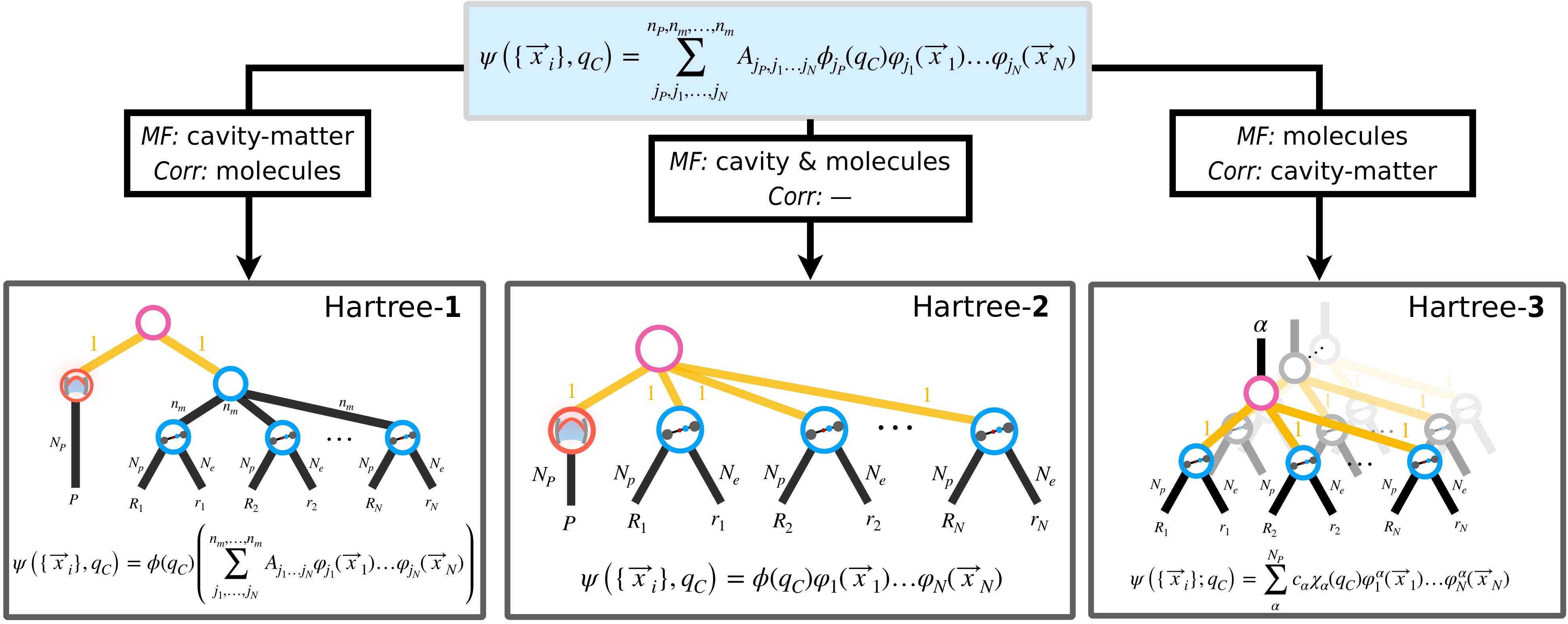}
    \caption{Mean-field approximations to the cavity-ensemble wavefunction $\psi(\lbrace\vec x_i\rbrace,q_C)$, $\vec x_i$ collects all DOFs of the $i$-th molecule. In the blue-shaded box, a correlated ansatz in MCTDH format is shown, in which $\psi(\lbrace\vec x_i\rbrace),q_C)$ is expanded as multi-configurational sum of Hartree products of single particle functions (SPF) of the cavity and molecular DOFs. The ansatz Hartree-1 employs a single SPF for the cavity and for the total matter wavefunction, the latter is then expanded in MCTDH format. This amounts to a mean-field (``MF'') description on the light-matter level while the ensemble wavefunction remains correlated (``Corr''). Hartree-2 restricts the MCTDH expansion to a single Hartree product, resulting in a full mean-field description of cavity and molecules. Hartree-3 includes cavity-matter correlation by expanding in $q_C$-dependent ensemble wavefunctions, where each ensemble wavefunction at the $\alpha$-th cavity DVR grid point is approximated by a single Hartree product. }
    \label{fig:Figure_6}
\end{figure*}

Earlier studies using the PFH have stressed the importance of intermolecular interactions induced by the DSE in determining local ground-state properties of molecules in the ensemble, especially at large per-molecule coupling strengths \cite{horak2025analytic,schnappinger2023cavity,philbin2023molecular}. This has motivated the use of a Hartree ansatz among molecular wavefunctions, which treats the two-body terms of the DSE at a mean-field level \cite{schnappinger2023cavity,sidler2024unraveling}. Of course, such a mean-field ansatz is only an approximation as it neglects intermolecular correlation. Hence, before turning to the collective strong coupling limit, which is experimentally feasible within a FP setup, we first study the validity of mean-field approximations to the cavity-ensemble wavefunction when transitioning from the weak to ultrastrong few-molecule coupling regime. Throughout Section \ref{sec:resC}, an ensemble of five molecules with fixed orientation is examined.

So far, all results presented in this study have been obtained with a converged correlated wavefunction ansatz. By exploiting the flexibility of the variationally-optimal tree-tensor network wavefunction we can construct approximate mean-field ansätze for the cavity-ensemble wavefunction and compare their performance to the fully-correlated calculations. Three mean-field approximations, shown schematically in Figure~\ref{fig:Figure_6}, are investigated: treating the interaction between cavity mode and the matter-subsystem as a whole on a mean-field level but retaining the intermolecular correlation (Hartree-1), mean-field among cavity and all molecules individually, i.e., also neglecting intermolecular correlation (Hartree-2), and mean-field among molecules but retaining full light-matter correlation (Hartree-3). 

Ansatz (Hartree-3)
\begin{align}
\psi(\lbrace{\vec x_i\rbrace},q_C)=\sum_{\alpha=1}^{N_P} \phi_\alpha(q_C)\prod_{j=1}^N \varphi^\alpha_j(\vec x_j),\label{eq:ansatz-h3}
\end{align}
with internal DOFs $\lbrace{\vec x_i\rbrace}$ of the $N$ molecules, and $N_P$ representing the number of configurations used to describe the system,
can be brought to treat correlations similarly to the cBOA-Hartree Fock (cBOA-HF) for molecular ensembles  \cite{schnappinger2023cavity}.
This is achieved by choosing the cavity wavefunctions as maximally localized orthogonal functions $\chi_\alpha(q_C)$ (DVR basis) around grid point $q_C^\alpha$. Specifically we take $\phi_\alpha(q_C)=c_\alpha\chi_\alpha(q_C)$,
where then $|c_\alpha|^2$ corresponds to the probability density at the cavity-displacement
$q_C=q_C^\alpha$. 
Now, by finding the optimal ground state for ansatz \eqref{eq:ansatz-h3} using the localized cavity basis, it is clear that one obtains
the optimal
molecular wavefunction at each cavity displacement, $\varphi_j^\alpha(\vec x_j)\to \varphi_j^\alpha(\vec x_j;q_C^\alpha)$.
Indeed, it is easy to see that removing the cavity DOF ``kinetic energy'' from the PFH and optimizing the total wavefunction, each product of molecular wavefunctions corresponds to the cBOA-HF ansatz, with the crucial difference that in cBOA-HF the nuclear coordinates are treated as parameters together with $q_C$. \cite{schnappinger2024molecular,schnappinger2023cavity}.
In our treatment with ansatz \eqref{eq:ansatz-h3}, where the second derivative cavity term is retained, and where both the cavity displacement and all molecular DOFs are quantum mechanical,
the level of theory corresponds
to the full Born-Oppenheimer approximation with second derivative non-adiabatic couplings included.

\begin{figure*}
    \centering
    \includegraphics[width=1.0\linewidth]{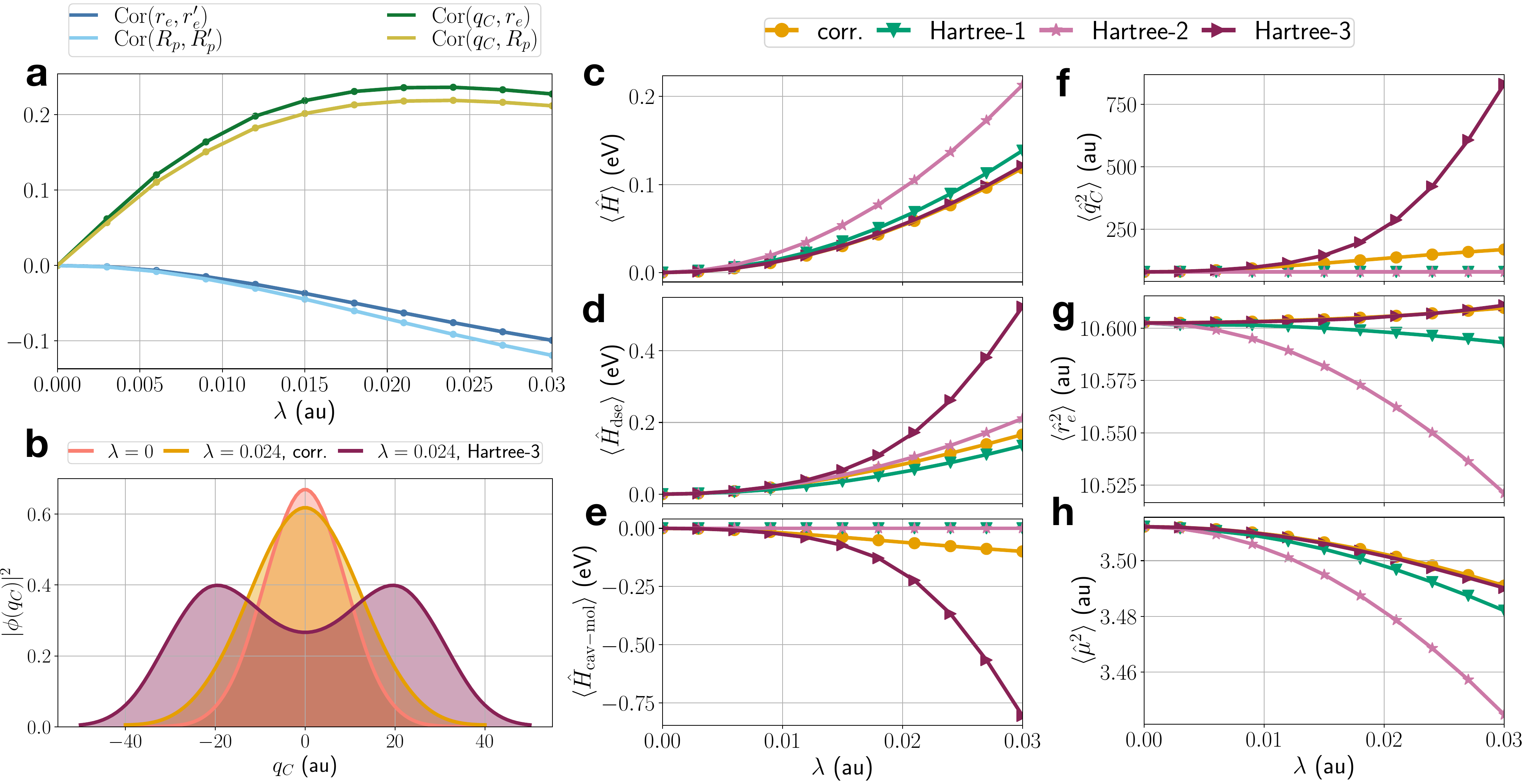}
    \caption{Analysis of correlation and mean-field approximations in the polaritonic ground state. All results are based on a fixed-orientation ensemble with five molecules and varying the per-molecule coupling strength $\lambda$. (a) Linear (Pearson) correlation coefficients, $\mathrm{Cor}(r_e,r_e')$ and $\mathrm{Cor}(R_p,R_p')$ denote correlation between electronic and nuclear DOFs of two different molecules from the ensemble, respectively. (b) One-dimensional reduced density along cavity mode displacement for correlated (``corr.'') wavefunction and Hartree-3 ansatz. (c)-(h) Comparison of mean-field approximations for computing observables as a function of per-molecule coupling strength.}
    \label{fig:Figure_7}
\end{figure*} 
We start by analyzing the correlations of the converged ground-state wavefunction in order to gauge their importance when transitioning from weak to ultrastrong coupling. In Figure~\ref{fig:Figure_7}a, the Pearson Correlation coefficient $\mathrm{Cor}(X,Y)=(\langle XY\rangle - \langle X\rangle\langle Y\rangle)/\sigma_X\sigma_Y$ is employed as a correlation measure between DOFs $X$ and $Y$, where $\sigma_X,\sigma_Y$ denote the standard deviations. The linear correlation $\mathrm{Cor}(X,Y)$ ranges from -1 to 1, where 1 indicates fully correlated DOFs, -1 indicates fully anti-correlated DOFs, and 0 indicates no correlation between them. On the one hand, the light-matter correlation, quantified by the linear correlations $\mathrm{Cor}(q_C,r_e)$ and $\mathrm{Cor}(q_C,R_p)$ increases quickly with $\lambda$ and remains the dominant correlation across the entire range of coupling strengths. On the other hand, intermolecular correlations between electrons and nuclei, $\mathrm{Cor}(r_e,r_e')$ and $\mathrm{Cor}(R_p,R_p')$, grow more slowly with $\lambda$ and become appreciable only around the onset of the USC regime. Thus, these correlations exactly reflect the transition from the one-body to many-body regime: up to $\lambda\approx0.01$ the interaction between individual molecules and cavity ($\hat H_{\mathrm{cav-mol}}$) and the one-body terms of $\hat H_{\mathrm{dse}}$ dominate. Beyond that, intermolecular correlations start to play an important role.

Despite the importance of light-matter correlation within the total ground-state wavefunction, a mean-field treatment on the light-matter level (Hartree-1) results in a good description of most observables for $\lambda<0.015$ in Figure~\ref{fig:Figure_7}c-h. Beyond the onset of the USC regime, deviations in local properties (e.g. $\langle \mu^2\rangle$), DSE and total energy become apparent. At the largest coupling strength, the relative errors with respect to the correlated reference solution are around 10\% for global energy contributions, and below 1\% for intramolecular properties shown in Figure~\ref{fig:Figure_7}c-h.
Nonetheless, Hartree-1 (and also Hartree-2) can neither describe modifications to the cavity mode fluctuations $\langle q_C^2\rangle$, nor the stabilization due to the bilinear interaction, which remains zero in these cases. This is because, within the light-matter mean-field, the cavity couples to the molecules only through their average dipole, factorizing the bilinear energy contributions as $\langle \hat q_C \hat \mu_j\rangle\approx \langle \hat q_C \rangle \langle\hat\mu_j\rangle$. The average dipoles $\langle\hat\mu_j\rangle$, however, must be zero in the ground state for all $\lambda$ due to inversion symmetry of the cavity SM Hamiltonian, such that $\langle\hat H_{\mathrm{cav-mol}}\rangle$ remains zero and the cavity mode fully decouples from the molecular ensemble. 

Thus, Hartree-1 and Hartree-2 properly account for
the DSE, but only in the one-body regime, where this term is well described by both mean-field approximations (Figure~\ref{fig:Figure_7}d). As discussed around Figure~\ref{fig:Figure3}), the local modifications to molecular structure are predominantly determined by the one-body part of the DSE, which can explain why both Hartree-1 and Hartree-2 result in rather small errors until the onset of USC. For larger coupling strengths, the one-body picture breaks down and, consequently, the full mean-field approximation (Hartree-2) becomes increasingly worse (Figure~\ref{fig:Figure_7}c-h).

Approximating the intermolecular dipole-dipole interactions mediated by the DSE as $\langle \hat\mu_i\hat \mu_j\rangle\approx \langle \hat\mu_i\rangle\langle \hat\mu_j\rangle$, discards this contribution since the total permanent dipole moment of the system is zero. The two-body DSE terms have a stabilizing effect on the ground state (Figure~\ref{fig:Figure3}) and can only grow through the build-up of intermolecular correlation. Therefore, Hartree-2 substantially overestimates the DSE and total energy with increasing coupling strength, at the largest coupling strength in Figure~\ref{fig:Figure_7}c-d ($\lambda=0.03$) by approximately 25\% and 75\% of the correlated value, respectively.
Furthermore, local modifications are overestimated too because rearrangement of charge densities within each molecule is the only way to minimize the energy in this scenario. Some of these shortcomings of a mean-field approximation to the entire cavity-ensemble wavefunction (Hartree-2) have also been observed in the context of cavity-coupled Bose-Einstein condensates \cite{alon2024entanglement}.

Let us now turn to the Hartree-3 ansatz, which employs an intermolecular mean-field approximation. cBOA and XF methods \cite{schnappinger2023cavity,flick2017atoms,hoffmann2018light} retrieve light-matter correlation, which is decisive for a correct description of the total ground state of the system, by making the molecular ensemble wavefunction cavity displacement-dependent. From this point of view, the results of Hartree-1 and Hartree-2 are electronic-nuclear ground-state solutions to the cBOA problem at zero cavity displacement since $\hat q_C$ is replaced by the constant expectation value $q_C=\langle \hat q_C\rangle=0$. Accordingly, Hartree-1 corresponds to a fully correlated solution whereas Hartree-2 is equivalent to ensemble cBOA-HF solution. Their pairwise comparison (Figure~\ref{fig:Figure_7}a-h, see also Figure~S7) shows that for $q_C=0$ numerically exact (i.e., correlated) and mean-field solution quickly deviate with increasing $\lambda$. On the other hand, an analogous computation at a large cavity displacement $q_C=30$ (cf.~Figure~S7) results in excellent agreement between the mean-field and correlated solution within a cBOA framework. 

This can be explained by the fact that a intermolecular mean-field ansatz treats the uncorrelated part $\hat H_{\mathrm{uncorr}}=\sum_j \hat H_{\mathrm{mol}}^{(j)}-\lambda\omega_C q_c\cdot \sum_j\hat \mu_j + \frac{\lambda^2}{2}\sum_j \hat \mu_j^2$ of the ``clamped cavity displacement'' Hamiltonian exactly, and only approximates the two-body terms of the DSE. Since the second term grows linearly with $q_C$ and the DSE is independent of cavity operators, $\hat H_{\mathrm{uncorr}}$ dominates for sufficiently large $q_C$. At small cavity displacement $q_C\approx 0$, however, the DSE is the relevant energy contribution. When employing a mean-field ansatz in this region, the total energy is quickly overestimated with increasing $\lambda$ as the stabilizing, correlated terms $\langle \hat\mu_i\hat \mu_j\rangle$ are poorly treated (see above). Hence, since the cavity displacement changes the relative magnitude of uncorrelated and correlated energy contributions, the quality of the mean-field solution thus strongly depends on $q_C$ within the cBOA.

The global ground-state wavefunction obtained by minimizing the energy of the Hartree-3 ansatz can be seen as a weighted sum of such $q_C$-dependent mean-field solutions where the weights are given by the cavity wavefunction (cf.Eq.~\eqref{eq:ansatz-h3}). Overall, local molecular modifications and total energy agree well with the numerically exact reference calculations over the entire range of coupling strengths. Yet, large errors are found for the cavity mode fluctuation, DSE and bilinear interaction at the onset of the USC regime which are significantly larger than the errors of Hartree-1 and Hartree-2. 

This can be traced back to the breakdown of the Hartree product ansatz around $q_C=0$ for increasing $\lambda$: the overestimation of energy in this region decreases the weight of these solutions during minimization, shifting it to larger displacements where the mean-field solution becomes accurate. This is clearly reflected in the comparison of cavity displacement densities $|\phi(q_C)|^2$ (Figure~\ref{fig:Figure_7}b) for a high coupling strength $\lambda=0.024$. While the correlated solution has a maximum at $q_C=0$ with increased variance compared to 
the uncoupled case, the mean-field ansatz leads to two maxima away from $q_C=0$ at $q_C\approx \pm20$. With increasing $\lambda$ these maxima move further apart, resulting in a steeper increase of $\langle q_C^2\rangle$ compared to the correlated ansatz. This further results in a severe overestimation of $\langle\hat{H}_{\mathrm{cav-mol}}\rangle$, since the wavefunction resides preferentially at large displacements, and of $\hat{H}_{\mathrm{dse}}$ due to the lack of stabilizing two-body contributions. At the largest coupling strength, $\langle\hat H_{\mathrm{dse}}\rangle$ and $\langle\hat{H}_{\mathrm{cav-mol}}\rangle$ are overestimated by factors of about 2 and 8, respectively (Figure~\ref{fig:Figure_7}d-e).

We have seen that a globally reliable and accurate description of the ground state of the PFH for strong light-matter coupling requires a wavefunction ansatz which includes both light-matter and intermolecular correlations. Specifically, the cavity mode fluctuations are very sensitive to correlations within the system. Through light–matter interactions, the cavity responds to the entire molecular ensemble. In doing so, it simultaneously probes all intermolecular correlations induced by the all-to-all coupling in the DSE. Likewise, the bilinear term energy contribution and the total polarization of the ensemble, as given by the DSE, include the interactions between all constituents of the system. Hence, we refer to them as \emph{global} properties. Conversely, when computing \emph{local}, i.e., intramolecular properties of a specific molecule, the remaining ensemble and cavity mode are traced out. We conclude that mean-field methods can lead to large errors in global ground-state properties of the PFH, while local properties are accurately described if light-matter correlation is treated properly.

We stress that the errors of the Hartree-3 ansatz are entirely due to the lack of intermolecular correlation: for $N=1$ the reference calculation and Hartree-3 coincide (Figure~S8), while for larger $N$ agreement can be reached if each $q_C$-dependent ensemble wavefunction $\psi_{\alpha}(\vec X)$ is expanded in a multi-configurational sum instead of a single Hartree product (Figure~S9).
Moreover, the break-down of the Hartree-3 ansatz not solely depends on the per-molecule coupling strength $\lambda$ but also on the number of molecules. As shown in Figure~S10, the intermolecular mean-field solution for $N=10$ becomes unstable at lower coupling strengths compared to $N=5$, below the onset of USC.

\subsection{Taking the large-$N$ limit}\label{sec:largeN}
\begin{figure*}
    \centering
    \includegraphics[width=\linewidth]{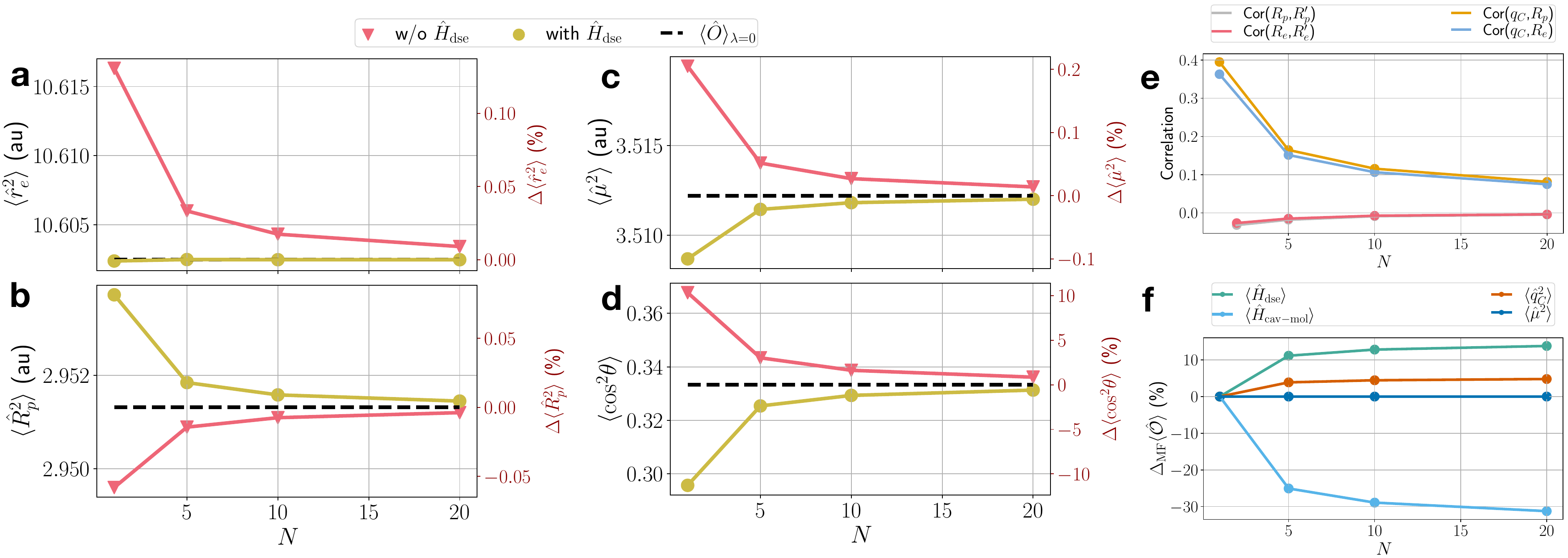}
    \caption{Ground-state modifications and correlations in the thermodynamic (i.e., large-$N$ limit). The collective coupling strength is kept constant at 0.02\,au in all cases by employing a scaled per-molecule coupling strength $\lambda_{\mathrm{eff}}=0.02/\sqrt{N}$. The effect on single-molecule observables (a)-(d) and correlations  (e) is shown. Single-molecule observables are given in absolute terms (left axis) and relative to the bare-molecule value, $\Delta \langle\mathcal{\hat O}\rangle = (\langle\mathcal{\hat O}\rangle_{\lambda} - \langle\mathcal{\hat O}\rangle_{\lambda=0})/\langle\mathcal{\hat O}\rangle_{\lambda=0}$ (right axis, in dark-red). (f) Relative error of the intermolecular mean-field approximation (Hartree-3, cf.~Figure~\ref{fig:Figure_6}) with respect to the converged correlated wavefunction, $\Delta_{\mathrm{MF}}\langle \hat{\mathcal O} \rangle=(\langle\mathcal{\hat O}\rangle_{\mathrm{MF}} - \langle\mathcal{\hat O}\rangle_{\mathrm{corr}})/|{\langle\mathcal{\hat O}\rangle_{\mathrm{corr}}|}$. In (a)-(d) freely rotating molecules are considered, whereas in (e) molecules are fixed to be fully aligned with the cavity-mode polarization axis.}
    \label{fig:Figure5}
\end{figure*}
Finally, we investigate how cavity-induced correlations and changes to the structure of individual molecules behave under collective strong coupling. To this end, we fix the collective coupling strength and converge to the thermodynamic limit by employing the effective per-molecule coupling strength $\lambda_{\mathrm{eff}}=0.02/\sqrt{N}$, as introduced in the beginning.

Our numerically exact, first-principles calculations indicate a quick decay of local cavity-induced modifications with growing ensemble size in Figure~\ref{fig:Figure5}a-d and Figure~S6a-c. Crucially, this suggests a vanishing impact of the cavity on individual molecular observables in the ground state under collective VSC. Calculations with and without DSE result in small ($\langle \hat r_e^2\rangle$) to large ($\langle \mathrm{cos}^2\theta\rangle$) modifications of the local observables for $N=1$, but already agree closely with the out-of-cavity value at $N=20$. In the large-$N$ limit, the discrepancies between the PFH and mDH disappear, as both converge to the out-of-cavity value, albeit from opposite directions.

 These findings can be rationalized by a perturbative analysis similar to that in Refs.~ \cite{philbin2023molecular} and  \cite{fischer2024cavity} (see Appendix for details) which treats the light-matter interaction as a perturbation to the matter Hamiltonian. For a local observable $\langle\hat{\mathcal{O}}_j\rangle$ which only acts on DOFs of the $j$-th molecule, the leading order corrections due to coupling to the cavity originate from the one-body part of $\hat H_{\mathrm{dse}}$ and the interaction between the $j$-th molecule and cavity.
 Both contributions scale with the square of the per-molecule coupling strength and solely depend on zeroth-order (i.e., field-free) properties of a single molecule. Consequently, they decay when taking the large-$N$ limit by scaling $\lambda$ with $1/\sqrt{N}$ and $N\rightarrow\infty$, i.e., the $\lambda^2$-scaling of the perturbative result in the Appendix implies a $1/N$-dependence of the local modifications. When plotting the results of Figure~\ref{fig:Figure5} against $1/N$ (see Figure~S11) we find a linear behavior, confirming the $1/N$-dependence.
 The onset of a $N$-dependence of single-molecule observables and the transition from a quadratic to linear scaling with $\lambda$ in Figure~\ref{fig:Figure2} and S3 marks the breakdown of the perturbative expressions. However, this occurs only at per-molecule coupling strengths which are experimentally unfeasible \cite{de2025there}.

Intramolecular properties, light-matter, and pairwise intermolecular correlations decay when taking the large-$N$ limit, with the latter quickly approaching zero (Figure~\ref{fig:Figure5}e). This is in agreement with previous studies \cite{galego2015cavity}. Interestingly, although pairs of molecules become increasingly uncorrelated in the large-$N$ limit, this does not imply that a mean-field approximation to the ensemble wavefunction (Hartree-3) becomes exact in this limit. While intramolecular properties such as $\langle \mu^2\rangle$ are treated correctly, errors in the cavity energy contributions $\langle \hat H_{\mathrm{dse}}\rangle$ and $\langle \hat H_{\mathrm{cav-mol}}\rangle$ increase with $N$ and saturate around 12\% and 30\%, respectively. The same holds for the cavity ground-state fluctuations $\langle q_C^2\rangle$ where the mean-field error saturates at approximately 5\% (Figure~\ref{fig:Figure5}e).

This is in line with the observation of the previous sections that intermolecular correlations are critical for a correct description of the global properties of the ground state, although at a local intramolecular scale their impact is small. Moreover, the intermolecular mean-field approximation was found to break down at lower per-molecule couplings when the ensemble size was increased. This suggests that the quality of the mean-field ansatz deteriorates with increasing per-molecule coupling or number of molecules.  

In the perturbative treatment of single-molecule observables, only one-body terms contribute. Correlated wavefunction corrections arising from the two-body part of the dipole self-energy (DSE) do not appear up to second order in $\lambda$ (cf.~\eqref{eq:pert-local} in Appendix). This insensitivity explains why local observables are well captured by a mean-field approximation, as already observed for the few-molecule strong coupling case in Section~\ref{sec:changes}. Locally, the one-body term scales with $\lambda^2$, such that modifications to local observables vanish regardless of how intermolecular correlations are treated.

Yet, from a global viewpoint, the one-body part scales with $N\lambda^2$ whereas the two-body interactions (due to the all-to-all nature of the DSE) with $N(N-1)\lambda^2$. Thus, the relative importance of the correlated part of the DSE increases with the ensemble size. In particular, when taking the thermodynamic limit by replacing $\lambda$ by $\lambda_{\mathrm{eff}}=\lambda/\sqrt{N}$ and letting $N\rightarrow\infty$, the pairwise molecular interactions decay with $1/N$, in agreement with the pairwise intermolecular correlations $\mathrm{Cor}(r_e,r_e')$ and $\mathrm{Cor}(R_p,R_p')$ in Figure~\ref{fig:Figure5}e. Their contribution to the total DSE, on the other hand, does not vanish and saturates with increasing $N$. Similarly, the cavity mode which couples to the entire molecular ensemble (and vice versa), also depends sensitively on the inter-molecular correlations.

A similar argument can be found in Ref.~\cite{sidler2024collectively} to justify the importance a self-consistent treatment of the DSE in the thermodynamic limit. Importantly, our numerically exact simulations and perturbation theory show that the intermolecular correlations induced by the DSE do not affect the local molecular ground-state properties in the large-$N$ limit, which simply approach the bare molecular value. Notwithstanding, for global energy contributions of the PFH and for the cavity mode they are relevant. Accurately capturing these observables in the large-$N$ limit requires a multi-configurational self-consistent treatment (cf.~Figure~\ref{fig:Figure5}f).

\section{Discussion}

We have performed first-principle simulations of the polaritonic ground state of SM molecule ensembles, with full-quantum treatment of all degrees of freedom, including the photonic, electronic, nuclear and rotational degrees of freedom, across a wide parameter range. Our simulations employed the full PFH, including counter-rotating terms and dipole-self energy. Such non-perturbative, numerically exact simulations of molecular
ensembles in FP cavities have been elusive so far, but are accessible within the tree-tensor network-based ML-MCTDH method. Besides analytical solutions and the plethora of simulations based on more approximate models and techniques, they present an important building block for guiding future theoretical efforts toward a microscopic understanding of polaritonic chemistry. 

In this work, we applied this approach to the ground state in a few-molecule strong coupling setting, and subsequently converged to the ``thermodynamic'' (i.e., large-$N$) collective strong-coupling limit by systematically increasing the number of molecules and rescaling of the coupling strength by $1/\sqrt{N}$.
In the few-molecule strong coupling scenario, we find that increasing the per-molecule coupling strength results in a response of rotational, vibrational and electronic properties. While this response is pronounced for rotation and --to a smaller degree-- vibrations, modifications to the electronic structure are marginal, even when each molecule is ultrastrongly coupled to the cavity. These modifications scale with the per-molecule coupling, not the collective coupling strength until the onset of the single-molecule USC regime. 

In this range of couplings, both cavity-induced energy contributions, the DSE and bilinear term, are extensive quantities, i.e., their per-molecule value is independent of the ensemble size, due to dominating one-body contributions in DSE and bilinear term. Since the balance between DSE and bilinear term eventually shapes the modifications of local molecular ground-state properties, they also are independent of the ensemble size. For even larger coupling strengths, this picture breaks down due to the buildup of intermolecular correlations which results in collective (ensemble-size dependent) scaling of global energy contributions. Local modifications, on the other hand, were found to be rather insensitive to the presence of intermolecular correlations. In the gas phase, reorientation away from the cavity axis strongly limits the cavity-induced modifications to other molecular DOFs and the appearance of any collective effects. 

Moreover, our simulations can confirm the importance of the DSE in the USC regime to obtain a stable ground-state solution. Nonetheless, it is important to recall that the per-molecule coupling strengths required to observe local ground-state modifications and instability are much larger than realizable values in FP cavities. As demonstrated recently \cite{de2025there}, reaching the single- or few-molecule regime is not possible through interaction with a purely transverse electromagnetic field but requires longitudinal charge-charge interactions between molecules and a material structure (e.g. nanoparticle). A quantized description of such an experimentally realizable setting would, however, require an \emph{ab initio} Hamiltonian different from the PFH \eqref{eq:full-ham} which -- most importantly -- would not contain the DSE. First-principles simulations of an adequate \emph{ab initio} Hamiltonian which includes a quantized description of the ``cavity'' material \cite{medina2021few,jamshidi2023coupling} are possible with the methodology of this paper, but are beyond the scope of this study. We rather note that connecting simulations (of this study and other publications) at thus large per-molecule couplings based on the PFH \eqref{eq:full-ham} to vibropolaritonic chemistry in FP cavities must be approached with utmost care. 

Taking the large-$N$ limit resembles more closely the experimental settings in the field of VSC-modified ground-state chemistry. When increasing the number of molecules while keeping the collective coupling (proportional to the Rabi splitting) constant, our first principles simulations show that local molecular modifications quickly decay. This implies that cavity-induced changes to the molecular ground state become minute under typical experimental conditions which involve $10^6-10^{12}$ molecules -- even in the \textit{collective} ultrastrong coupling regime. In addition to that, we find that these results are not specific to VSC. As shown in Figure~S6d-f, the vanishing impact of the cavity on local molecular observables extends to the ground state under ESC as well. 

A perturbative analysis, treating the coupling strength $\lambda$ as the perturbative parameter, is in line with these results. A $\lambda^2$-dependence of local cavity-induced modifications is obtained, which implies a $1/N$-decay of cavity-induced ground-state modifications when considering collective strong coupling. Consequently, the amply discussed ``large-$N$ problem'' \cite{schwennicke2024molecular,martinez2018polariton,martinez2019triplet} applies to local ground-state modifications too.

Previous theoretical works have already pointed out the absence of a collective scaling of local ground-state observables under collective (ultra)strong coupling. For instance, equilibrium bond lengths and molecular re-aligment are found to scale with the per-molecule coupling strength in \cite{galego2015cavity} and \cite{cwik2016excitonic} as well as the potential energy curve of an individual molecule within the ensemble in \cite{martinez2018can,li2020origin}. A minuscule internuclear correlation in the collective strong coupling limit has already been predicted in \cite{galego2015cavity}. 
These works were based on the mDH (omiting the DSE), expansion in field-free electronic states, as well as perturbative treatment. Our study aimed at numerically exact solution of the full coupled electron-nuclear-photonic problem of Eq.~\eqref{eq:full-ham}, reaching the same conclusions.%

In addition to that, our results agree well with recent experiments. In Ref.~\cite{hoblos2025does}, the authors found that the equilibrium between high-spin and low-spin ground state 
of a spin-crossover molecule remained unchanged, even for collective coupling strengths close to the USC regime. Probing molecules under collective VSC with nuclear magnetic resonance in Ref.~\cite{patrahau2024direct} showed no measurable effects of the cavity on the molecular electronic structure.

Special attention should be paid to the role of correlations in the ground-state cavity-ensemble wavefunction. Light-matter correlations are important for an accurate description of local properties of molecules, cavity displacement and global energy contributions, especially when entering the per-molecule USC regime. Neglecting intermolecular correlations in the wavefunction ansatz via a mean-field approximation, as in the cBOA-HF method, can introduce substantial errors in global properties. In contrast, local molecular modifications are captured accurately, reflecting their insensitivity to intermolecular correlation.

This trend persists when taking the large-$N$ limit, although the correlation between pairs of molecules rapidly approaches zero with increasing ensemble size. This manifests the importance of a correlated ensemble wavefunction ansatz for a globally correct description of the cavity-ensemble ground state, both in the per-molecule and collective (ultra)strong coupling regime.
Furthermore, the quality of a cBOA-HF ansatz depends strongly on the cavity displacement, where small cavity displacement result in larger errors.

 In conclusion, fully converged \emph{ab initio} simulations using the Pauli–Fierz Hamiltonian show that, in the thermodynamic limit, they agree with the results of the molecular Dicke Hamiltonian. At the same time, local molecular properties remain essentially unchanged in this limit.
 In particular,
 our results show that
 \emph{there is no substantial modification to local molecular properties without each separate molecule being in the USC regime}.
 Together, these findings demonstrate that, for most practical applications in polaritonic chemistry, quantum mechanical descriptions of ensembles can be accurately carried out with a molecular Dicke–type Hamiltonian using field-free molecular states as a basis.
 In future work, our laboratory will apply the methodology presented in this letter to investigate earlier theoretical predictions on time-dependent processes in excited states that were based on more approximate methods. This is expected to provide valuable insight for the polaritonic chemistry community.

\section*{Acknowledgments}
The authors acknowledge support by the state of Baden-Württemberg through bwHPC
and the German Research Foundation (DFG) through grant INST 35/1597-1 FUGG.
O.V. acknowledges support through the German Research Foundation (DFG)
Collaborative Research Center 1249.

\bibliography{shinmetiu}

\section{Appendix}
In the following, we derive a perturbative expression for the expectation value of a general local operator $\hat{\mathcal{O}}_j$ which only acts on the DOFs of the $j$-th molecule.
To this end, we treat $\hat H_{\mathrm{cav-mol}}+\hat{H}_{\mathrm{dse}}=-\omega_C \hat q_C\vec\lambda\cdot\hat{\vec D} + \frac{1}{2}\left(\vec\lambda\cdot\hat{\vec D}\right)^2$ as the perturbation and expand the unperturbed system in a product basis of photon number states and eigenstates of the molecular Hamiltonian, $|n_C;\phi_m^{(l)}\rangle=|n_C\rangle\otimes|\phi_m^{(l)}\rangle$. Here, $|n_C\rangle$ denotes a photon number state with $n$ photons in the cavity, and  $|\phi_m^{(l)}\rangle$ the $m$-th rovibronic eigenstate of the $l$-th molecule with all other molecules in their absolute ground states. In intermediate normalization, the first-order corrected wavefunction reads
{\small
\begin{eqnarray}
    |\psi^{(1)}\rangle= &&\lambda\sqrt{\frac{\omega_C}{2}}\sum_{l=1}^N\sum_{n\neq0}\frac{\langle\phi_n|\hat\mu|\phi_0\rangle}{\omega_C+\omega_n}|1_C;\phi_n^{(l)}\rangle\label{eq:cav-m-cont}\\
    &&-\frac{\lambda^2}{2}\sum_{l=1}^N\sum_{n\neq0}\frac{\langle\phi_n|\hat\mu^2|\phi_0\rangle}{\omega_n}|0_C;\phi_n^{(l)}\rangle\label{eq:dse1-cont}\\
     &&-\frac{\lambda^2}{2}\sum_{l\neq k}^N\sum_{n,m\neq0}\frac{\langle\phi_n|\hat\mu|\phi_0\rangle\langle\phi_m|\hat\mu|\phi_0\rangle}{\omega_n+\omega_m}|0_C;\phi_n^{(l)}\phi_m^{(k)}\rangle\label{eq:dse2-cont},
\end{eqnarray}}
where contribution \eqref{eq:cav-m-cont} originates from the bilinear cavity-molecule interaction and \eqref{eq:dse1-cont}-\eqref{eq:dse2-cont} from the dipole-self energy. For simplicity of notation $\hat\mu$ already denotes the polarization axis-projected dipole operator, i.e., $\hat\mu=(\hat R_p - \hat r_e)\cos\theta$. The amplitudes in \eqref{eq:cav-m-cont}-\eqref{eq:dse2-cont} depend on bare-molecule matrix elements of the dipole operator and squared dipole operator, as well as on the cavity mode energy $\omega_C$ and bare molecular excitation energies $\omega_n=\langle \phi_n|\hat H_{\mathrm{mol}}|\phi_n\rangle-\langle \phi_0|\hat H_{\mathrm{mol}}|\phi_0\rangle$. The molecular indices have been dropped in the amplitude expressions since all molecules have identical Hamiltonians in our model.

Plugging the first-order corrected wavefunction into the expression for the expectation value of $\hat{\mathcal{O}}_j$ yields
\begin{eqnarray}
    \langle\hat{\mathcal{O}}_j\rangle \approx &&\langle\psi^{(0)}|\hat{\mathcal{O}}_j|\psi^{(0)}\rangle + \langle\psi^{(1)}|\hat{\mathcal{O}}_j|\psi^{(0)}\rangle \nonumber\\&&+ \langle\psi^{(0)}|\hat{\mathcal{O}}_j|\psi^{(1)}\rangle + \langle\psi^{(1)}|\hat{\mathcal{O}}_j|\psi^{(1)}\rangle.
\end{eqnarray}
The unperturbed expectation value $\langle\psi^{(0)}|\hat{\mathcal{O}}_j|\psi^{(0)}\rangle$ is identical to the bare-molecular value $\langle \hat{\mathcal{O}}_j\rangle_{\lambda=0}$ in the main text. 
We only keep terms up to $O(\lambda^2)$ and thus obtain
{\small
\begin{eqnarray}
\langle\hat{\mathcal{O}}_j\rangle 
\approx && \mathcal{N}_\lambda^{-1}\bigg[ \left(1+(N-1)\lambda^2 A_{00}\right)\langle\hat{\mathcal{O}}_j\rangle_{\lambda=0}
 \nonumber\\
-&&\lambda^2 \sum_{n\neq 0}B_{n0}\langle\phi_n|\hat{\mathcal{O}}_j|\phi_0\rangle 
+ \lambda^2\sum_{n,m\neq 0}C_{nm}\langle\phi_n|\hat{\mathcal{O}}_j|\phi_m\rangle\bigg]
\label{eq:pert-local}
\end{eqnarray}}

with the appropriate normalization constant $\mathcal{N}_\lambda=1+\langle\psi^{(1)}|\psi^{(1)}\rangle$  \cite{sakurai2020modern}, which is given up to $O(\lambda^2)$ by
\begin{equation}
    \mathcal{N}_\lambda=1+N\lambda^2\frac{\omega_C}{2}\sum_{n\neq0}\frac{\langle\phi_0|\hat\mu|\phi_n\rangle\langle\phi_n|\hat\mu|\phi_0\rangle}{(\omega_C+\omega_n)^2}.
\end{equation}
and 
\begin{eqnarray}
    A_{00}=&& \frac{\omega_C}{2}\sum_{n\neq0}\frac{\langle\phi_0|\hat\mu|\phi_n\rangle\langle\phi_n|\hat\mu|\phi_0\rangle}{(\omega_C+\omega_n)^2},\\
    B_{n0}=&& \frac{\langle\phi_n|\hat\mu^2|\phi_0\rangle}{\omega_n}, \\
    C_{nm}=&& \frac{\omega_C}{2}\frac{\langle\phi_n|\hat\mu|\phi_0\rangle}{\omega_C+\omega_n}\frac{\langle\phi_0|\hat\mu|\phi_m\rangle}{\omega_C+\omega_m}.
\end{eqnarray}

From the expression \eqref{eq:pert-local} it becomes obvious that local ground-state modifications scale quadratically with the per-molecule coupling strength $\lambda$, and that they are an off-resonant effect due to the appearance of the sum, not the difference, $\omega_C+\omega_n$. Taking the large-$N$ limit of \eqref{eq:pert-local} by scaling $\lambda/\sqrt{N}$ and $N\rightarrow\infty$ results in $\langle\hat{\mathcal{O}}_j\rangle \rightarrow \langle\hat{\mathcal{O}}_j\rangle_{\lambda=0}$.

\end{document}